\documentclass[graybox, envcountchap]{svmult}

 \usepackage[numbers]{natbib}

\usepackage{mathptmx}        
\usepackage{amsmath}
\usepackage{amssymb}
\usepackage{color}
\usepackage{helvet}          
\usepackage{courier}         
\usepackage{dirtree}

\usepackage{makeidx}        
\usepackage{graphicx}        
\usepackage{subfig}

\usepackage{multicol}        
\usepackage[bottom]{footmisc}

\usepackage{hyperref}        
\hypersetup{colorlinks=true,urlcolor=blue}

\usepackage[misc]{ifsym}

\DeclareRobustCommand{\VAN}[3]{#2}
\let\VANthebibliography\thebibliography
\def\thebibliography{\DeclareRobustCommand{\VAN}[3]{##3}\VANthebibliography}
\newcommand{\der}{\text{d}}
\newcommand{\rhoc}{\rho_\text{c}}
\newcommand{\YSZ}{Y_{\text{SZ}}}
\newcommand{\YX}{Y_{\text{X}}}
\newcommand{\LX}{L_{\text{X}}}

\newcommand{\Mpc}{\;{\rm Mpc}}
\newcommand{\hMpc}{{\ifmmode{\;h^{-1}{\rm Mpc}}\else{$h^{-1}$Mpc}\fi}}
\newcommand{\hkpc}{{\ifmmode{\;h^{-1}{\rm kpc}}\else{$h^{-1}$kpc}\fi}}
\newcommand{\hMsun}{{\ifmmode{\;h^{-1}{\rm {M_{\odot}}}}\else{$h^{-1}{\rm{M_{\odot}}}$}\fi}}
\newcommand{\Msun}{\;\rm {M_{\odot}}}
\newcommand{\kelvin}{\;\rm K}
\newcommand{\LxUnits}{\;\rm erg \text{ } s^{-1}}
\newcommand{\Mstar}{{\ifmmode{\;M_{*}}\else{$M_{*}$}\fi}}
\newcommand{\Mhalo}{{\ifmmode{\,M_{\rm halo}}\else{$M_{\rm halo}$}\fi}}
\newcommand{\ltsima}{$\; \buildrel < \over \sim \;$}
\newcommand{\gtsima}{$\; \buildrel > \over \sim \;$}
\newcommand{\lsim}{\lower.5ex\hbox{\ltsima}}
\newcommand{\gsim}{\lower.5ex\hbox{\gtsima}}

\makeindex             

\begin{document}


\title{Machine Learning applications to Galaxy Clusters}
\author{Gustavo Yepes  and  Daniel de Andrés}
\institute{Gustavo Yepes (\Letter) \at Departamento de Física Teórica and CIAFF, Universidad Autónoma de Madrid, Cantoblanco 28049, Madrid, Spain,  \email{gustavo.yepes@uam.es}
\and Daniel de Andrés  \at Nonlinear Dynamics, Chaos and Complex Systems Group, 
Departamento de Geología, Física y Química Inorgánica, Universidad Rey Juan Carlos, 
Tulipán s/n, 28933 Móstoles, Madrid, Spain  \email{daniel.deandres@urjc.es}}
%
%
\maketitle

\abstract{This chapter reviews the application of  Artificial Intelligence (AI) techniques to the study of galaxy clusters, covering both theoretical developments and their use as tools to infer cluster properties from a variety of observational tracers. We discuss recent advances in mass estimation from SZ, X-ray, optical, and dynamical data, highlighting the ability of AI  methods to capture non-linear features, projection effects, and complex cluster morphologies beyond more classical  approaches. In addition, we present other emerging applications, including the emulation of baryonic physics from N-body simulations, the characterization of dynamical states and mergers, and the analysis of the diffuse components such as the intracluster light. Particular emphasis is placed on the role of simulations in training these models, the impact of baryonic modelling, and the need for a robust uncertainty quantification and interpretability. Finally, we outline current limitations and future prospects, stressing the importance of combining flexible simulation strategies with AI  techniques to fully exploit next-generation surveys for precision cosmology.}


\section{Introduction}

Galaxy clusters, as the one depicted  in Fig \ref{fig:thebullet},  represent the final stage of hierarchical structure formation, arising from the non-linear growth of primordial density perturbations. They are the most massive gravitationally bound systems in the Universe, with typical masses of $10^{14}$--$10^{15} \Msun$ and characteristic sizes of $\sim 2 \Mpc$ \citep{kravtsov2012}. Early dynamical studies, most notably by Zwicky, showed that masses inferred from galaxy velocity dispersions exceed the luminous (stellar) content by roughly two orders of magnitude, providing the first observational evidence for dark matter \citep{Zwicky1933}.

The gravitational potential of galaxy clusters is dominated by dark matter, and clusters can be described as systems composed of three main components:

\begin{figure}
    \centering
    \includegraphics[width=1\linewidth]{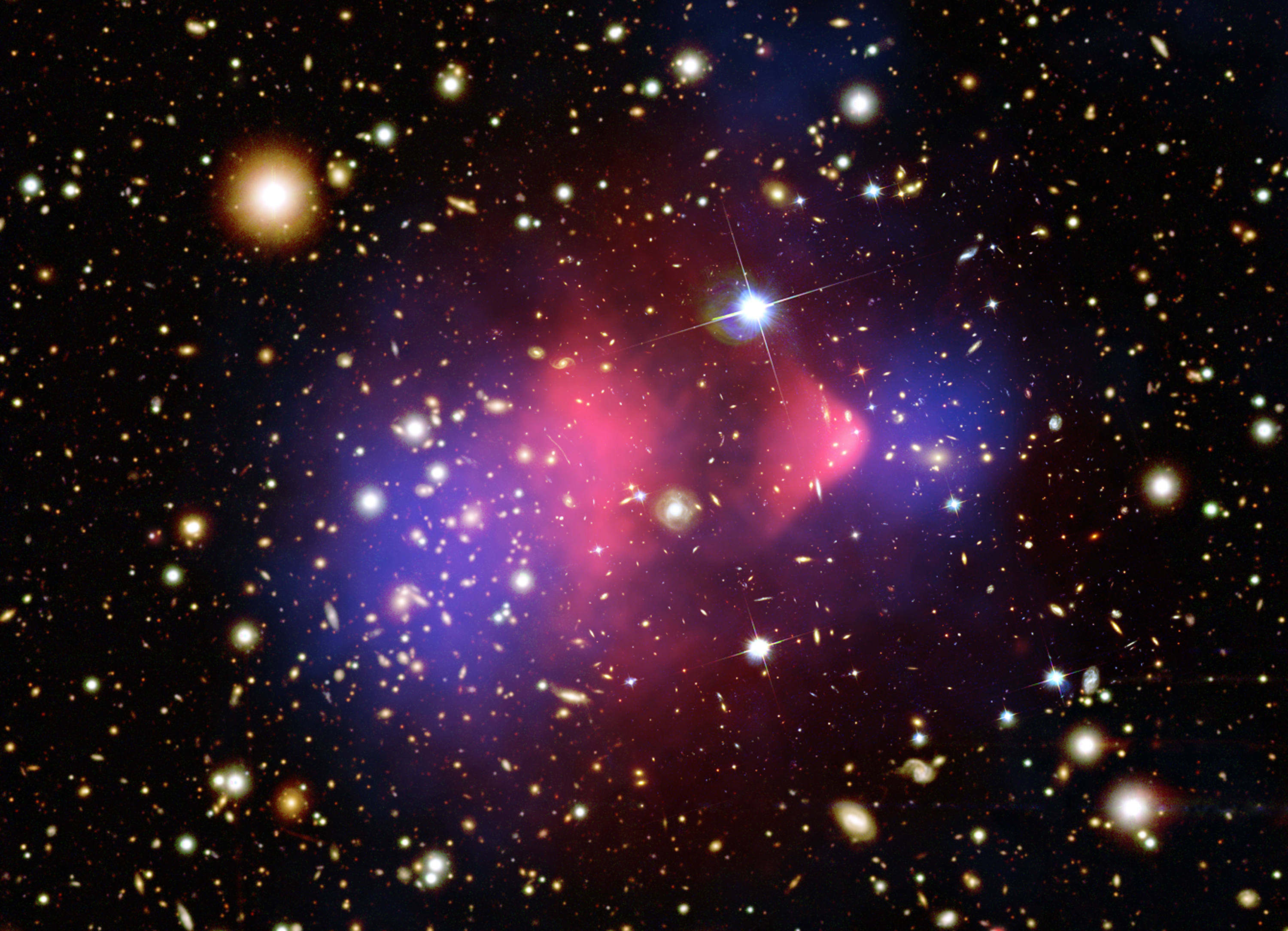}
    \caption{The ``bullet cluster", galaxy cluster 1E 0657-556. The two pink clumps correspond to the hot gas detected in X-rays, and the optical image from the Magellan and the Hubble Space Telescope shows the galaxies in orange and white. The blue area corresponds to the concentration of mass inferred by gravitational lensing effects. (Credit: X-ray: NASA/CXC/CfA/M. Markevitch et al.; optical: NASA/STScI, Magellan/U. Arizona/D. Clowe et al.; lensing map: NASA/STScI ESO WFI, Magellan/U. Arizona/D. Clowe et al. Public Domain.) }
    \label{fig:thebullet}
\end{figure}

\begin{enumerate}

\item \textbf{Dark matter} constitutes the dominant mass component ($\sim 80$\%). Although not directly observable, it is inferred through its gravitational effects, such as lensing and galaxy dynamics. Its radial distribution is commonly described by the Navarro–Frenk–White (NFW) profile \citep{NFW}:
\begin{equation}
\rho_{\text{DM}}(r)=\rho_{s}\left[
\left(\frac{r}{r_s}\right)\left(1+\frac{r}{r_s}\right)
\right]^{-1}.
\end{equation}

\item \textbf{Galaxies} contribute a subdominant fraction of the total mass ($\sim 5$--$10$\%) and are primarily observed at optical and near-infrared wavelengths. Clusters typically host a central Brightest Cluster Galaxy (BCG), often a massive cD galaxy. 
Large galaxy surveys such as the Sloan Digital Sky Survey (SDSS; \cite{York2000SDSS}), the Dark Energy Spectroscopic Instrument (DESI; \cite{DESI2016}), and wide-field photometric surveys like the Dark Energy Survey (DES; \cite{DESCollaboration2016}), the Legacy Survey of Space and Time (LSST) at the Vera C. Rubin Observatory \citep{Ivezic2019LSST}, \textit{Euclid} \citep{Laureijs2011Euclid}, and the \textit{Nancy Grace Roman Space Telescope} \citep{Spergel2015Roman}, together with deep targeted observations from \textit{JWST} \citep{Gardner2006JWST}, have significantly improved, and will further enhance, the characterization of galaxy cluster populations across cosmic time.

\item \textbf{The intracluster medium (ICM)} accounts for $\sim 10$--$15$\% of the total mass and consists of a hot, diffuse plasma with temperatures $T \sim 10^{7}$--$10^{8},\kelvin$. It emits primarily in X-rays, with luminosities up to $L_{X}\sim 10^{45},\LxUnits$, and is observed with instruments such as \textit{Chandra}, \textit{XMM-Newton} \citep{chexmatesurvey}, and \textit{eROSITA} \citep{erositasurvey}. The ICM also produces the Sunyaev--Zel'dovich (SZ) effect \citep{Sunyaev1972}, arising from inverse Compton scattering of CMB photons by hot electrons, which induces spectral distortions at the level of $10^{-4}$--$10^{-5}$. This effect is observed at millimetre wavelengths by facilities such as \textit{Planck} \citep{PlanckPSZ2}, \textit{SPT} \citep{SPTcatalog}, \textit{ACT} \citep{ACTcatalog}, and \textit{NIKA2} \citep{Mayet2020:Nika2}.

\end{enumerate}

\begin{figure}
    \centering
    \includegraphics[width=0.8\linewidth]{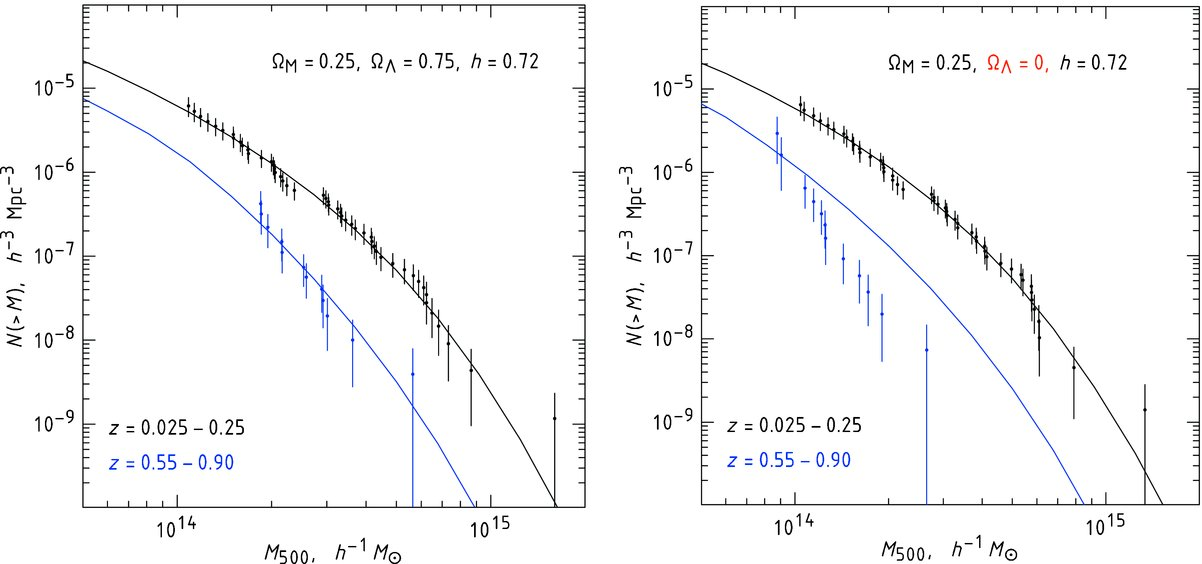}
    \caption{Illustration of the sensitivity of the cluster mass function to the underlying cosmological model. The data correspond to the mass estimates of X-ray selected clusters in the Chandra Cluster Cosmology Project. Image taken from \citet{Nclusters}}
    \label{fig:counts}
\end{figure}

Therefore, galaxy clusters provide a fundamental laboratory for studying both the formation of large-scale structure and the complex astrophysical processes operating in dense environments. In addition, their abundance as a function of mass and redshift constitutes one of the most powerful cosmological probes to constrain the matter content of the Universe (see Fig \ref{fig:counts}) . However, cluster mass is not a direct observable. It must be inferred from the emission of their different components across a wide range of wavelengths, from millimetre to X-ray bands. 

To establish robust relations between the total mass of galaxy clusters and observable quantities, it is essential to model consistently their three main components: dark matter and baryons in the form of gas and stars. Only through such modelling can one calibrate the link between the underlying mass and the multi-wavelength observables of the baryonic components, which ultimately allows cluster masses to be inferred from observations. In this context, the formation and evolution of galaxies and galaxy clusters require cosmological simulations that include not only gravitational dynamics but also the relevant baryonic processes, such as gas hydrodynamics, star formation, stellar feedback, and AGN feedback.

A major challenge in cosmological simulations is the trade-off between computational cost, simulated volume, and numerical resolution. Ideally, one would like to model large cosmological volumes with sufficient resolution to capture both the large-scale structure of the Universe and the internal properties of dark matter haloes. However, this goal is particularly demanding in the case of galaxy clusters.  Since they are the most  massive gravitationally bound systems,  they are intrinsically rare objects. Their low abundance implies that very large cosmological volumes are required in order to obtain statistically representative samples. At the same time, resolving the baryonic processes within cluster size halos, such as the formation of galaxies and the structure of the intra-cluster medium, demands very high resolution. These two requirements are in direct tension. As a result, full-volume hydrodynamical simulations that simultaneously achieve both large volumes and high resolution are computationally prohibitive, even with nowadays pre-exascale supercomputers. On the other hand, large N-body simulations can reach volumes of tens or   hundreds  of  Gpc$^3$, providing statistically robust samples of clusters, but they are limited to modelling only the dark matter component (see.  e.g. refs \cite{NBODYREV,Vogelsberger2020:reviewcosmosim} for recent reviews   on cosmological simulations).

A practical and widely adopted solution is provided by “zoom-in” simulations. In this approach, clusters are first identified in large, low-resolution cosmological volumes, ensuring a representative statistical sample. Selected objects are then re-simulated at much higher resolution, including full baryonic physics. This strategy allows one to retain both statistical relevance and detailed physical modelling, making it particularly well suited for studying galaxy clusters in detail.

In this context, simulation suites such as \textsc{The Three Hundred}\footnote{\url {https://www.nottingham.ac.uk/astronomy/The300/index.php}} project \citep{Cui2018,Cui2022}   provide an optimal framework, combining a large sample of clusters with detailed baryonic modelling and multiple physical implementations. This makes them especially valuable for  machine learning applications requiring  vast labelled datasets to be trained.

 In the last decade, machine learning (ML) and deep learning (DL) techniques have undergone an impressive  development that has also been applied to Astrophysical problems \citep{exmachina2023,Huertas2022}, enabling the identification of complex, non-linear relations in high-dimensional parameter spaces. These methods offer a natural framework to go beyond traditional approaches, which often rely on simplified scaling relations and assumptions such as hydrostatic equilibrium and spherical symmetry to connect cluster mass with observable quantities. In this sense, ML provides a powerful way to generalize the mappings between mass and multi-wavelength observables, as learned from numerical simulations, opening the path towards a more flexible and data-driven description of galaxy clusters.

In this context, we explore in this chapter the application of artificial intelligence models to derive the masses and other properties   of galaxy clusters. Such approaches are particularly well suited to exploit the richness and diversity of simulation results. However, ML methods—especially deep learning—require large, well-characterized labelled datasets for training, a requirement that only a limited number of simulation suites can currently fulfill.

\subsection{Classical methods for  mass estimation of galaxy clusters}\label{sec:massestimates}

The dynamical masses of galaxy clusters can be inferred under the assumption that the intra-cluster medium (ICM) is in \textbf{hydrostatic equilibrium (HE)} within the gravitational potential. This leads to:
\begin{equation}
M_{\text{HE}}(r) = -\frac{r^{2}}{G\mu m_p n_e (r)}\frac{\der P(r)}{\der r },
\end{equation}
where $P(r)$ can be derived from SZ observations, and $\mu$, $m_p$, and $n_e(r)$ denote the mean molecular weight, proton mass, and electron density profile, respectively. Alternatively, using X-ray observations \citep{Ansarifard2020}:
\begin{equation}
M_{\text{HE}}(r) = -\frac{rk_{B}T}{G\mu m_{p}}\left(
\frac{\der \ln n_e }{\der \ln r} +\frac{\der \ln T}{\der \ln r}
\right).
\end{equation}

In addition, \textbf{scaling relations} connect observable quantities with cluster mass \citep{Lovisari2022:scalinglaws}. The integrated Compton parameter is defined as $\YSZ \propto \int P_e \, dV$, i.e. the volume integral of the electron pressure, tracing the total thermal energy content of the ICM. Its X-ray analogue is defined as $\YX \equiv M_{\rm gas} T_X$, combining the gas mass and the spectroscopic gas temperature, and providing a low-scatter proxy of the same thermal energy. Under the self-similar assumption \citep{Kaiser1986:SS,Bohringer2012:self-similar}, observables such as $\YSZ$,  $\YX$ and the total bolometric  luminosity emitted in X-ray band,  $\LX^{\text{bol}}$,  scale with mass and redshift as:
\begin{equation}
\YSZ \sim E(z)^{2/3}M^{5/3}, \quad
\LX^{\text{bol}} \sim E(z)^{7/3}M^{4/3}, \quad
\YX \sim E(z)^{2/3}M^{5/3}.
\end{equation}
These integrated quantities are defined within a sphere of radius $R_{\Delta}$ corresponding to an overdensity $\Delta$ relative to the critical density:
\begin{equation}
\rhoc (z) = \frac{3H^{2}(z)}{8\pi G},
\end{equation}
leading to:
\begin{equation}
M_{\Delta} = \Delta \rhoc (z) \frac{4\pi}{3} R_{\Delta}^{3}.
\label{eq:moverdens}
\end{equation}
Different values of the overdensity limit  $\Delta=2500,500,200$ are often used in literature. 

The SZ signal is described by the Compton-$y$ parameter:
\begin{equation}
y = \frac{\sigma_{\text{T}}k_{\text{B}}}{m_{\text{e}}c^{2}}\int n_{\text{e}}T_{\text{e}}dl,
\end{equation}
which implies $\YSZ \sim M_{\text{gas}}T_e$. Similarly, the X-ray emissivity is:
\begin{equation}
\epsilon(\nu) = n_e n_p \Lambda(T, Z,\nu),
\end{equation}
from which $\LX$ is obtained.

In practice, deviations from self-similarity arise due to non-gravitational processes such as radiative cooling, AGN feedback, and mergers. Therefore, scaling relations are generalized as:
\begin{equation}
\YSZ= A_{\text{SZ}} E(z)^{\alpha_{\text{SZ}}}M^{\beta_{\text{SZ}}},
\end{equation}
\begin{equation}
\LX^{\text{bol}}= A_{\text{bol}} E(z)^{\alpha_{\text{bol}}}M^{\beta_{\text{bol}}},
\end{equation}
\begin{equation}
\YX= A_{\text{X}} E(z)^{\alpha_{\text{X}}}M^{\beta_{\text{X}}}.
\end{equation}

However, the calibration of these relations remains uncertain, leading to systematic discrepancies between simulations and observations \citep{Planck2013:SZClusterCounts}. This directly impacts cluster mass estimates and, consequently, cosmological constraints.

The mass of galaxy clusters can also be estimated by measuring the velocity dispersion $\sigma_v$ of the galaxy members. The mass estimated by this velocity dispersion is named as the dynamical mass of the galaxy cluster, and typically scales by a factor
\begin{equation}
    \sigma_v \sim   M^{1/3}   \text{ ,}
\end{equation}
which is the factor found in N-body simulations \citep{evrard2008virial} and consistent with the virial theorem. The above equation is typically generalised as in \cite{Ntampaka2015}:
\begin{equation}
    \log (\sigma_v)  = \alpha \log(M)+\beta \text{,}
\end{equation}
where $\alpha$ and $\beta$ are free parameters. However, the $M–\sigma_v$
relation includes oversimplifications
such as gravitational equilibrium, perfect selection of the member galaxies and spherical symmetry \citep{Old2018}.

\subsection{Calibration of the mass bias} \label{sec:mass_bias}

Mass estimates based on HE are known to be biased. This bias is defined as:
\begin{equation}
b = \frac{M_{\text{tot}}-M_{\text{HE}}}{M_{\text{tot}}},
\end{equation}
where $M_{\text{tot}}$ is the true mass. Typical values from simulations and weak lensing (WL) studies suggest $1-b \sim 0.8$--$0.9$, while CMB-based analyses require lower values ($1-b \sim 0.6$), leading to a well-known tension.

Weak Lensing   measurements provide independent mass estimates without assumptions on the ICM, generally supporting lower bias values \citep{Herbonnet2020:WL}. Similarly, simulations such as {\sc The Three Hundred} predict biases $b \sim 0.1$--$0.2$ \citep{Gianfagna2023}.

On this regard, as we will mention later, our  works using machine learning techniques applied to simulated datasets \citep{deAndres2022Planck} have provided alternative, independent,  calibrations consistent with simulation-based estimates. Overall, reconciling these different approaches remains a key challenge, although recent eROSITA results suggest a possible resolution of this tension \citep{eROSITAcounts}.

\subsection{Structure of the chapter}
As we mentioned in the abstract, the main goal of this chapter is to review the applications of novel Machine Learning and Deep Learning techniques to infer the physical properties of galaxy clusters, with an special emphasis on the determination of the total mass of these objects from a variety of observational tracers ( see \S \ref{massinference}).  Other applications  of ML and DL to galaxy clusters, such as the emulation of baryonic properties in DM only simulations, the dynamical  mass accretion history, the intra cluster light, etc, will be reviewed in \S \ref{others}. We will finish the chapter with the summary and conclusions in \S \ref{summary}.

\section{Cluster Mass inference from AI techniques}\label{massinference}

Machine Learning  techniques have recently emerged as a robust alternative to the traditional mass estimation methods for galaxy clusters described in \ref{sec:massestimates}, which are affected by both systematic biases and significant intrinsic scatter. In contrast, ML approaches are designed to learn the mapping between observables and total mass directly from simulations, enabling the capture of non-linear dependencies, projection effects, and complex morphological features that are typically not incorporated in standard analyses.

\subsection{Mass inference from SZ emission}

As shown in numerical simulations, the integrated SZ signal is tightly correlated with cluster mass \citep{Cui2018}. The main advantage of ML-based approaches is their ability to exploit the full information content of the signal—including morphology and dynamical state—rather than relying solely on integrated quantities within a fixed  overdensity radius, e.g. $R_{500}$ (see Eq. \ref{eq:moverdens})

The application of deep learning to the SZ effect was first explored by \cite{Gupta2020}, where the mResUNet architecture is trained on mock  microwave sky maps from the \textsc{Magneticum} simulations\footnote{\url { https://www.magneticum.org/}}  that include CMB anisotropies, dusty and radio galaxies, instrumental noise, and the cluster’s own SZ signal. The model was  able to infer cluster masses with a $1\sigma$ uncertainty of $\sim 20$\%.

\begin{figure}
    \centering
    \includegraphics[width=0.6\linewidth]{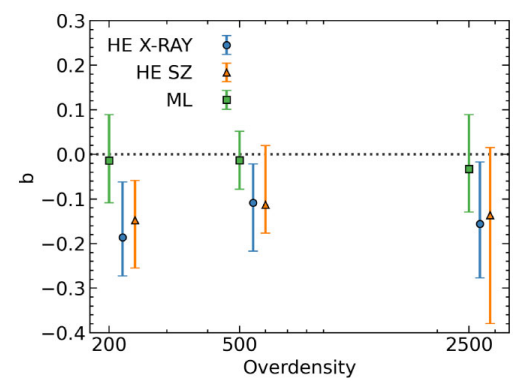}
    \caption{The comparison of the mass bias $b$ as a function of the overdensity (200,500 and 2500) between the methods based on Hydrostatic Equilibrium hypothesis  and Autoencoder+Random Forest. Image taken from  \cite{Ferragamo2023}.}
    \label{fig:Ferragamo2023}
\end{figure}

A  hybrid deep learning approach combining a Convolutional Autoencoder with a Random Forest was  used by in  \cite{Ferragamo2023}  to recover,  not only integrated masses,  but also the full spherical average  3D mass profiles. The method was  trained on mock observations derived from the {\sc The Three Hundred} simulations. A key result of this work is the significant reduction in scatter of the reconstructed profiles compared to classical parametric methods, indicating that ML techniques can effectively capture departures from self-similarity and complex cluster morphologies. As shown in  Fig. \ref{fig:Ferragamo2023}, this approach outperforms the standard hydrostatic equilibrium (HE) method in both bias and scatter.

In \cite{deAndres2022Planck}, deep learning techniques were applied to real CMB data for the first time, demonstrating that these methods are not limited to idealized or proof-of-concept studies. In this work, Convolutional Neural Networks (CNNs) trained on mock Compton-$y$ maps from the {\sc The Three Hundred} simulations were applied to \textit{Planck} observations.

The model is able to predict cluster masses directly from SZ maps, achieving a performance comparable to—or exceeding—that of the standard $\YSZ$–$M$ scaling relation, while avoiding explicit assumptions about the thermodynamic state of the intracluster medium. Importantly, the method naturally incorporates morphological information and projection effects, which are typically neglected in traditional analyses.

\begin{figure}
    \centering
    \includegraphics[width=0.8\linewidth]{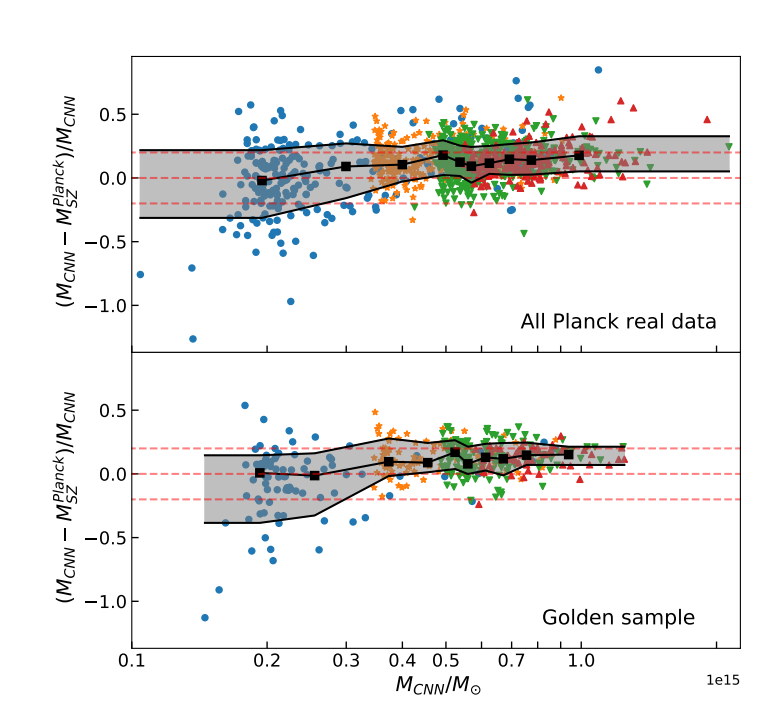}
    \caption{The mass bias  between the CNN inferred mass and the original  mass estimated by the Planck collaboration for  galaxy clusters detected by the Planck satellite. Image taken from  ref \cite{deAndres2022Planck}. }
    \label{fig:deAndres2022Planck}
\end{figure}

Moreover, this approach provides an independent calibration of the hydrostatic mass bias (see Fig.~\ref{fig:deAndres2022Planck}), yielding values consistent with expectations from numerical simulations. A key implication of this result is that the reliability of the training simulations is critical: the inferred masses are conditioned on the underlying physical modelling implemented in the simulations, which constitutes one of the main limitations of the method.

A major outcome of this work is the release of a CNN-based full-sky cluster mass catalogue\footnote{\url{https://github.com/The300th/DeepPlanck}}.

\subsection{Mass inference from X-ray emission}

Deep learning models have also demonstrated a strong capability to improve cluster mass estimates from X-ray observables. In \cite{Ntampaka2019}, CNN architectures were used for the first time to infer cluster masses directly from  X-ray photon-count images computed from the  Illustris TNG simulations\footnote{\url {https://www.tng-project.org/}}
achieving a reduction in scatter to $\sim 8$\%, compared to $\sim 18$\%  for standard techniques.

In \cite{Ho2023}, a realistic mock dataset of \textit{eROSITA} observations derived from the {\sc Magneticum} simulations incorporating instrumental effects and contamination from AGN  was used as training.  By combining multiple X-ray energy bands (low, medium, and high) within a CNN framework, the study systematically assessed the performance of deep learning models in recovering cluster masses from X-ray data  achieving  a mass scatter of $\sim 16$\%, significantly improving over traditional X-ray scaling relations and demonstrating that incorporating multi-band X-ray information enhances the robustness and accuracy of mass inference.

More recently, \cite{Krippendorf2024} developed a CNN-based framework to infer cluster masses in the \textit{eROSITA} eFEDS field, demonstrating the applicability of these methods to survey-quality data.

\begin{figure}[h]
    \centering
    \includegraphics[width=1.0\linewidth]{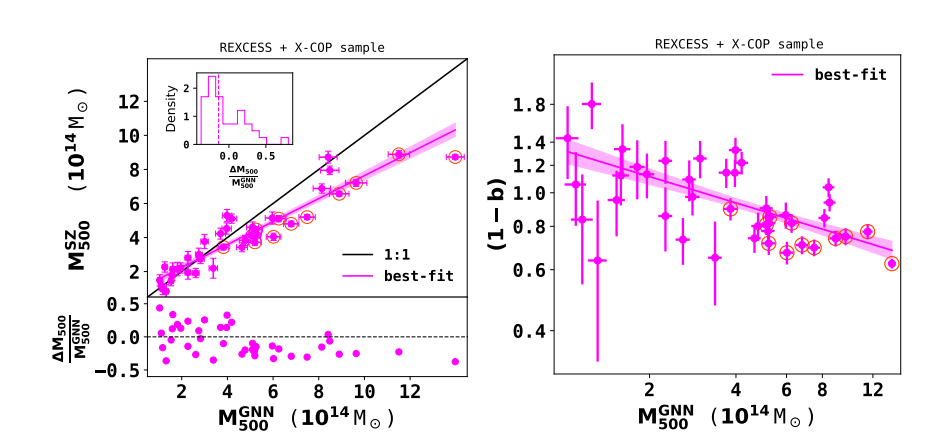}
    \caption{Comparison between GNN-inferred masses and SZ-based estimates: the left panel shows the mass–mass relation with residuals, while the right panel illustrates the mass-dependent SZ bias ($M_{500}^{\rm SZ}/M_{500}^{\rm GNN}$). The magenta line denotes the best fit with its $1\sigma$ uncertainty, and orange-circled points correspond to the X-COP sample. (Image taken from  \citet{Iqbal2025} )}
    \label{fig:Iqbal2025}
\end{figure}

Complementary to the different CNN approaches described above, \cite{Iqbal2025} introduced a Graph Neural Network (GNN) framework to estimate cluster masses from X-ray observations. The model was trained on {\sc The Three Hundred} simulations using radially sampled profiles of the ICM derived from X-ray data, achieving a dispersion of $\sim 7$\% on the test set. The trained network was subsequently applied to real datasets, including the XMM-Newton Cluster Structure Survey (REXCESS) and the XMM-Newton Cluster Outskirts Project (X-COP), as shown in Fig.~\ref{fig:Iqbal2025}.

\subsection{ Mass inference from galaxies }

Regarding dynamical mass estimates, these can be inferred from the full line-of-sight velocity distribution of cluster member galaxies, without requiring explicit assumptions about the dynamical state of the system. Early studies by \cite{Ntampaka2015,Ntampaka2016} employed the Support Distribution Machine (SDM; \citealt{SutherlandSDM}) algorithm—an extension of Support Vector Machines—to learn the mapping between velocity distributions and cluster mass. Compared to traditional $M$–$\sigma_v$ relations, they demonstrated that the scatter in the inferred masses can be reduced by approximately a factor of two using N-body simulations.

Subsequently, CNNs  were applied to regress cluster masses directly from phase-space representations (i.e. projected phase-space images of member galaxies) \citep{Ho2019}. These approaches further improved performance, reducing the scatter by $\sim 17$\%  and yielding more robust predictions than earlier ML methods. However, standard CNN frameworks do not natively provide uncertainty estimates; complementary approaches based on approximate Bayesian methods \citep{Ho2021} have been developed to infer masses with associated error bars.

Beyond standard CNNs, \cite{Kodi2020} demonstrated that Normalizing Flows methods  can be trained on simulated data (specifically on the Multidark\footnote{\url {https://www.cosmosim.org}}  simulations \cite {MDPL2}) and successfully applied to real clusters, including Coma, A1689, and several other massive systems. In a subsequent study, \cite{Kodi2021} implemented a simulation-based inference (SBI) framework using CNNs to estimate dynamical masses from the projected phase-space distribution, incorporating both sky positions and line-of-sight velocities. A notable application to real data is presented in \cite{Ho2022:ComaMass}, where deep learning methods were used to infer the mass of the Coma cluster, demonstrating that these techniques can be reliably extended beyond simulated datasets.

\subsection{Mass inference from multi tracer observations}

The use of multiple observables has also been explored within ML frameworks. \citet{Cohn2020} applied supervised learning algorithms (e.g. gradient boosting and random forests) to combine SZ, X-ray, and optical cluster observables, demonstrating a significant reduction in scatter relative to single-observable scaling relations. Their results indicate that multi-wavelength information helps break degeneracies inherent to individual mass proxies, leading to more robust and less biased mass estimates—an aspect that is particularly relevant for cosmological analyses, where systematic uncertainties in mass calibration directly propagate into parameter constraints. However, when exploiting the full spatial information in imaging data, combining multiple bands within a CNN framework does not necessarily yield substantial improvements; in particular, \cite{Yan2020} found that networks trained solely on Compton-$y$ maps can achieve higher precision than multi-channel inputs.

\begin{figure}
    \centering
    \includegraphics[width=1\linewidth]{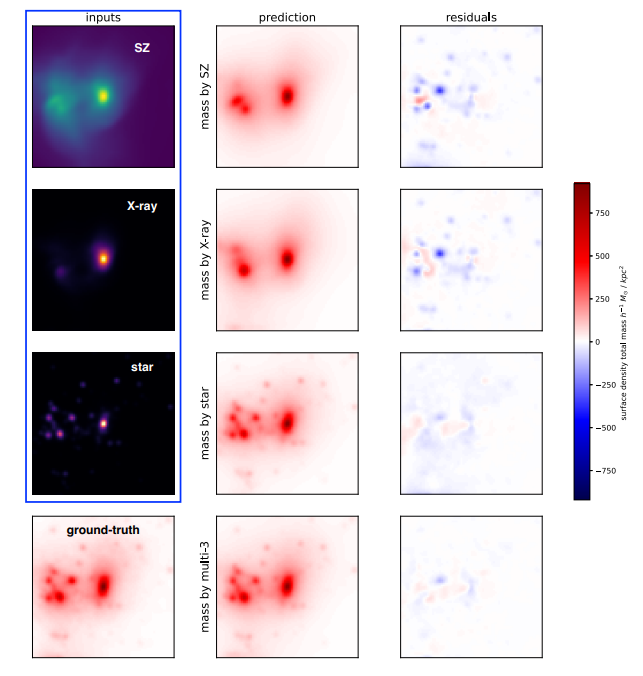}
    \caption{Mass map reconstruction from SZ, X-ray and stellar tracers. The residual maps (right column) corresponds to the difference between the predicted  and the ground-truth. Image  taken from  \citet{deAndres2024:MassMap}. }
    \label{fig:deAndres2024}
\end{figure}

Beyond estimating integrated masses within a given overdensity radius, \cite{deAndres2024:MassMap} advanced the field by using a U-Net architecture to reconstruct projected 2D total mass maps from SZ, X-ray, and stellar components. Their results show that combining optical and gas tracers significantly improves performance, reducing the scatter in derived integrated quantities by a factor of $\sim 2$ (see Fig.~\ref{fig:deAndres2024}). In addition, a closely related approach is presented in \cite{Benyas2024ApJ}, where score-based generative models are used to infer projected gas and dark matter density maps of galaxy clusters from mock SZ and X-ray observations.

Finally,  ML and DL  have also been widely applied to gravitational lensing analyses in galaxy clusters, both in the weak and strong regimes. In weak lensing, CNNs  have been used to infer cluster masses directly from shear maps, capturing non-linear features beyond standard two-point statistics \citep[e.g.][]{Gupta2018,Fluri2019}. In the strong lensing regime, DL  models have also been developed to accelerate lens modelling and to reconstruct mass distributions from lensed images, bypassing traditional parametric approaches \citep[e.g.][]{Hezaveh2017,PerreaultLevasseur2017}.  While these methods constitute powerful data-driven approaches for cluster mass inference, a detailed discussion is beyond the scope of this work, as dedicated chapters in this volume specifically address machine learning applications to gravitational lensing techniques.

Overall, a consistent picture emerges from these studies: ML and DL methods systematically outperform traditional scaling relations in terms of scatter and bias, particularly when spatial information is included. While classical estimators reduce the complexity of cluster observables to a small number of integrated quantities (e.g., $\LX$, $\YSZ$, $\YX$), ML approaches exploit the full information content of the data, including morphology, substructure, and projection effects. This leads to more accurate and robust mass estimates, with typical reductions in scatter by   a factor of $\sim 1.5$–2.

Nevertheless, these methods critically depend on the availability of large, well-characterized training datasets, which are typically provided by hydrodynamical simulations. In this context, the {\sc The Three Hundred} project constitutes a particularly suitable framework, as it combines a statistically representative sample of massive clusters with a diverse set of baryonic physics implementations, including different treatments of feedback and gas processes. This diversity is essential for capturing the range of astrophysical effects that impact observable–mass relations. Consequently, it has become a widely adopted benchmark for the development, training, and validation of ML-based mass estimators.

However, this reliance on simulations also introduces important limitations. ML models inherently learn the statistical and physical properties encoded in the training data, implying that any mismatch between simulations and real observations, such as inaccuracies in feedback prescriptions, selection effects, or unresolved physical processes, can lead to biased mass predictions. This issue, often referred to as the “simulation-to-reality gap,” highlights the need for careful validation against observational datasets, as well as for the development of domain adaptation techniques and hybrid approaches that combine simulations with real data.

In summary, ML and deep learning techniques provide a natural extension of classical methods, enabling a more flexible and physically informed mapping between observables and cluster mass. Their ability to capture complex, non-linear dependencies, incorporate high-dimensional data (e.g. images and phase-space information), and account for morphological and dynamical features makes them particularly well suited for next-generation surveys such as \textit{eROSITA}, \textit{Euclid}, and future  CMB experiments. In this context, controlling systematic uncertainties in cluster mass calibration is a key requirement for precision cosmology, and ML-based approaches offer a promising pathway to achieve this goal, provided that their limitations and underlying assumptions are carefully understood and quantified.

However, several important challenges remain. One of the most prominent is the interpretability of deep learning models. In this context, a consistent result across the literature for both X-ray and SZ analyses \citep{Ntampaka2019,Yan2020,deAndres2022Planck,Ho2023} is that interpretability techniques, such as saliency maps \citep{Simonyan2013,Selvaraju2017GradCAM} and feature-visualization methods like Google's DeepDream \citep{Mordvintsev2015}, tend to highlight pixels in the outskirts of clusters rather than in their central regions. This indicates that the features driving mass inference are predominantly located at intermediate and large radii.   Such behaviour is physically motivated, as cluster cores are strongly affected by complex baryonic processes (e.g. cooling, AGN feedback), which introduce additional scatter and are less tightly correlated with total mass, as also seen in hydrodynamical simulations \citep{Cui2018,Cui2021}. In contrast, optical-based analyses \citep{Yan2020} show that the most relevant information is associated with the distribution of member galaxies, reflecting the direct connection between galaxy dynamics and the underlying gravitational potential.

Another major challenge of DL models  concerns  the estimation of uncertainty. Most existing approaches rely on approximations to a full Bayesian treatment, which remains computationally intractable for deep neural networks.  SBI methods \citep{Kodi2021,deAndres2022Planck} provide one such framework, where the posterior distribution is approximated from simulated data by modelling the likelihood of summary statistics, often via techniques such as kernel density estimation. Alternatively, approximate variational inference methods \citep{Ho2021} model the posterior over network weights using a parameterized variational distribution, enabling the propagation of uncertainties into the inferred masses. Despite these advances, robust and well-calibrated uncertainty estimates still remain an open problem.

A further limitation is the lack of standardized benchmarking. A wide variety of models, ranging from CNNs to GNNs and normalizing flows, have been proposed for cluster mass estimation, often targeting similar observables and datasets. However, these methods are rarely evaluated on common benchmark datasets with consistent metrics, making it difficult to perform fair comparisons and to assess their relative performance and generalization capabilities. Establishing standardized benchmarks and validation protocols is therefore essential for the systematic development of ML-based approaches in cluster cosmology.

\section{Other applications of AI to galaxy clusters}\label{others}

Although mass inference has been the primary focus of deep learning applications to galaxy clusters, a growing number of studies have demonstrated that AI techniques can also address a broader range of astrophysical problems. These include the emulation of baryonic effects, the characterization of dynamical states, and the extraction of diffuse components such as intracluster light, among others.

A particularly active area is the emulation of baryonic properties from N-body simulations. Since full hydrodynamical simulations are computationally expensive, several works have explored the use of machine learning and deep learning to map dark-matter-only simulations into realistic baryonic fields. For instance, \cite{Rothschild2022}, \cite{deandres2022baryon}, and \cite{Caro2025} developed models that predict gas, stellar, and thermodynamic properties of galaxy clusters directly from N-body inputs. These approaches significantly reduce computational cost while retaining accuracy, enabling the generation of large mock catalogues required for survey analyses. In this context, ML-based emulators provide an efficient alternative to traditional subgrid modelling, while also allowing for fast exploration of parameter space. 

Symbolic regression has been used to enhance scaling relations between mass and observables \citep{Wadekar2023PNAS}. More generally, machine-learning methods provide an efficient means of discovering novel scaling relations with reduced scatter.

Beyond emulation, AI methods have been applied to characterize the dynamical and environmental properties of clusters and their galaxy populations. \cite{Contreras2023ML} investigated the identification and evolution of galaxy pairs using ML techniques, providing insight into interaction-driven processes within dense environments. Similarly, \cite{Srivastava2025} used Random Forest and CNNs to infer halo formation times, demonstrating that non-linear features in the BCG and in the IntraCluster Light (ICL, the  diffuse stellar component bound to the gravitational potential of the cluster),   encode information about assembly history that is not captured by standard methods. These results highlight the potential of ML to probe secondary halo properties and assembly bias.

AI has also started to be used to study transient and dynamical events such as cluster mergers. In \cite{Arendt2024}, CNNs  were applied to simulated SZ and X-ray maps to identify and classify merging systems, making use of the morphological and thermodynamical features present in the data. This type of approach is particularly useful for linking merger activity with observable signatures and for building cluster samples selected by dynamical state.

\begin{figure}
    \centering
    \includegraphics[width=0.8\linewidth]{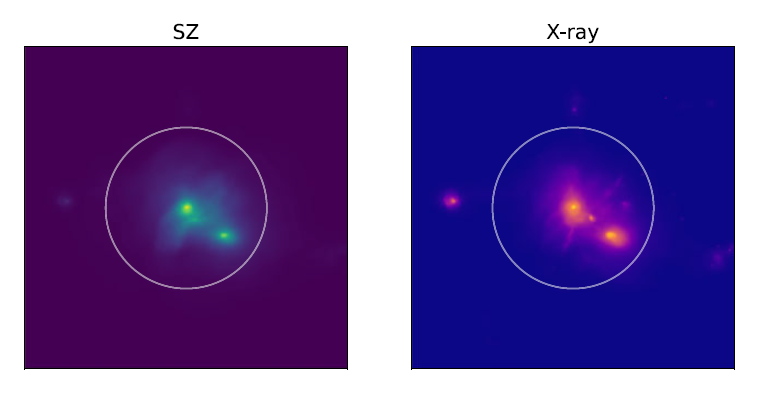}
    \caption{An example of SZ and X-ray images of merging clusters. White circles correspond to  $R_{200}$. The  CNN method developed by \cite{Arendt2024} is able to classify these events as ``merger".}
    \label{fig:placeholder}
\end{figure}

\begin{figure}
    \centering
    \includegraphics[width=1\linewidth]{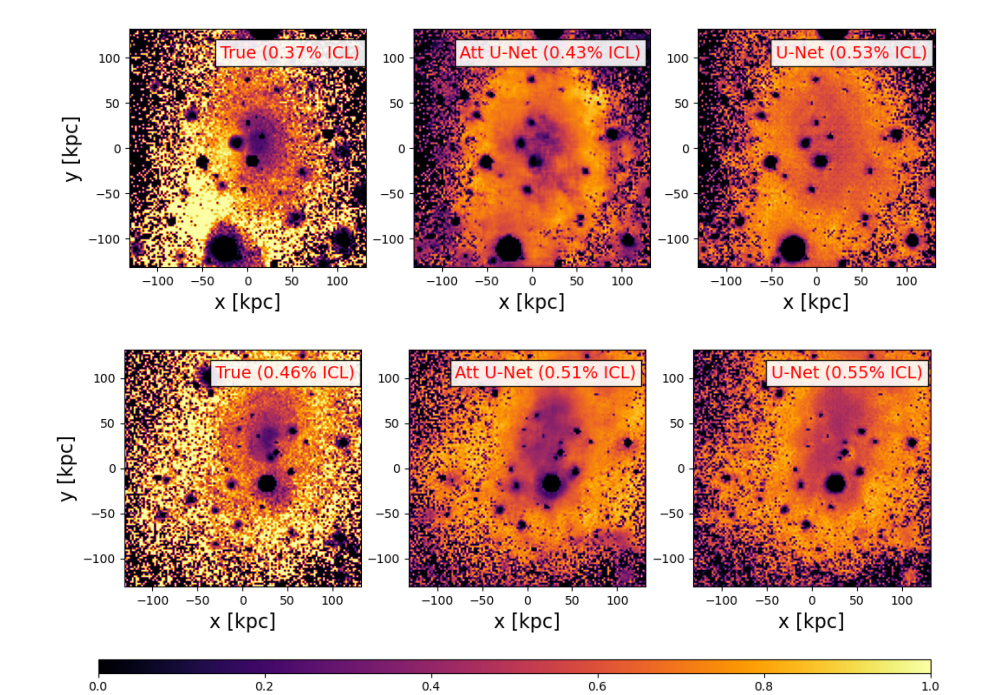}
    \caption{Comparison of the pixel-by-pixel  ICL mass fraction for two randomly selected  halos from the test set (left panels), and the  Attention U-Net (centre), and U-Net (right) model predictions. The colour map indicates the ICL fraction within each image, with the total ICL fraction reported in parentheses. This Figure is taken from ref \cite{Marini2024}}
    \label{fig:marini2024}
\end{figure}

Another relevant application concerns the study of the ICL.
In \cite{Marini2022,Marini2024}, deep learning methods based on  U-Net-like architectures, were employed to detect and reconstruct the ICL component. These models are trained on hydrodynamical simulations, where the ICL can be explicitly defined at the star  particle level, enabling a supervised segmentation of the diffuse stellar component and a detailed recovery of its spatial distribution. The ability of deep learning methods in predicting the ICL can be visualized in  Fig. \ref{fig:marini2024}, taken from ref  \citep{Marini2024}).

A complementary strategy is presented in \cite{Canepa2025}, where a supervised deep learning framework based on a convolutional encoder (ResNet-like architecture) is developed to estimate the ICL fraction directly from imaging data. In contrast to simulation-based approaches, the model is trained on large samples of synthetic images constructed from real Hyper Suprime-Cam observations, where artificial ICL profiles are injected to mimic the diffuse component. The training is then transferred to real data through a fine-tuning stage on a smaller sample of observed clusters. Rather than performing explicit segmentation, the model predicts the ICL fraction as a global property of the system, encoding the output as a probabilistic distribution.

These two approaches highlight complementary directions: simulation-based methods provide physically grounded labels and enable spatially resolved reconstructions, while observation-driven synthetic training offers a scalable path to process large survey datasets with minimal assumptions. Since the ICL is closely linked to the assembly history of clusters, these developments open new avenues to probe hierarchical structure formation in a statistically robust manner.


Overall, these applications illustrate that AI techniques extend well beyond mass estimation, offering powerful tools to model baryonic physics, characterize dynamical states, and extract faint signals. As observational datasets continue to grow in size and complexity, such methods are expected to play an increasingly important role in fully exploiting the information content of galaxy cluster surveys.

\section{Summary and Conclusions}\label{summary}

In this review, we have presented a comprehensive overview of the application of artificial intelligence (AI) techniques to galaxy clusters, with particular emphasis on mass inference and related problems. Machine learning (ML) and deep learning (DL) methods have demonstrated a clear advantage over traditional approaches by exploiting the full information content of observational data, including non-linear features, projection effects, and complex morphological signatures. These capabilities translate into improved precision and reduced biases in mass estimates across multiple observables, such as SZ, X-ray, optical, and dynamical tracers.

Beyond mass estimation, AI techniques are increasingly being applied to a broader range of problems, including baryonic field emulation, dynamical state classification, merger identification, and the characterization of diffuse components such as the intracluster light. These developments highlight the versatility of ML methods and their potential to extract additional physical information from large and complex datasets, which will be essential for upcoming surveys.

Despite these advances, several challenges remain. In particular, the dependence of ML models on simulations introduces systematic uncertainties associated with baryonic physics. Since different feedback prescriptions and subgrid models can significantly alter observable–mass relations, it is essential to properly marginalize over these uncertainties when training and applying ML models. In this context, approaches inspired by the {\sc CAMELS}\footnote{\url{https://www.camel-simulations.org/}}  project \citep{CAMELS_presentation}, where a wide range of baryonic and cosmological parameters are systematically varied, provide a promising framework. Extending this strategy to the cluster regime is a natural next step for the field.

However, generating statistically significant samples of massive clusters requires very large cosmological volumes, making full-box hydrodynamical simulations computationally prohibitive. In this regard, the strategy adopted by {\sc The Three Hundred} project, based on multiple high-resolution zoom-in simulations of cluster regions, offers a particularly efficient and flexible alternative. This approach enables the re-simulation of the same Lagrangian regions under different baryonic physics models and numerical resolutions, facilitating the  construction  of  controlled training datasets. Such flexibility is crucial for building ML models that are robust to uncertainties in baryonic modelling and capable of generalizing to real observations.

In conclusion, AI methods are becoming a fundamental component of galaxy cluster studies, providing powerful tools to improve mass calibration and to explore new physical diagnostics. Their successful application to precision cosmology will, however, require careful treatment of systematic effects, particularly those associated with baryonic physics, as well as the development of simulation strategies tailored to the needs of ML training.

\begin{acknowledgement}
The authors acknowledge the \textsc{The Three Hundred} collaboration for their substantial efforts in producing one of the most comprehensive simulated galaxy cluster datasets, which provides an excellent benchmark for training AI models. The simulations were carried out on the MareNostrum supercomputer at the Barcelona Supercomputing Center, with computational resources provided by the Red Española de Supercomputación.

 This work was supported by the Agencia Estatal de Investigación (AEI, Spain)  under project PID2024-156100NB-C21, funded by MCIN/AEI/10.13039/501100011033.

\end{acknowledgement}

\label{sec:1}

\bibliographystyle{mnras}
\bibliography{ml_bibtex}

@ARTICLE{NBODYREV,
       author = {{Angulo}, Raul E. and {Hahn}, Oliver},
        title = "{Large-scale dark matter simulations}",
      journal = {Living Reviews in Computational Astrophysics},
         year = 2022,
        month = dec,
       volume = {8},
       number = {1},
          eid = {1},
        pages = {1},
          doi = {10.1007/s41115-021-00013-z},
archivePrefix = {arXiv},
       eprint = {2112.05165},
 primaryClass = {astro-ph.CO},
       adsurl = {https://ui.adsabs.harvard.edu/abs/2022LRCA....8....1A}
}

@article{Gupta2018,
  author = {Gupta, Arushi and Lanusse, François and Mandelbaum, Rachel and Ribli, Dezső and Li, Yin and Szalay, Alexander S.},
  title = {Non-Gaussian information from weak lensing data via deep learning},
  journal = {Physical Review D},
  year = {2018},
  volume = {97},
  number = {10},
  pages = {103515},
  doi = {10.1103/PhysRevD.97.103515}
}

@article{Fluri2019,
  author = {Fluri, Janis and Kacprzak, Tomasz and Lucchi, Aurelien and Refregier, Alexandre and Hofmann, Thomas},
  title = {Cosmological constraints from noisy convergence maps through deep learning},
  journal = {Physical Review D},
  year = {2019},
  volume = {100},
  number = {6},
  pages = {063514},
  doi = {10.1103/PhysRevD.100.063514}
}

@article{Hezaveh2017,
  author = {Hezaveh, Yashar D. and Levasseur, Laurence Perreault and Marshall, Phil J.},
  title = {Fast automated analysis of strong gravitational lenses with convolutional neural networks},
  journal = {Nature},
  year = {2017},
  volume = {548},
  pages = {555--557},
  doi = {10.1038/nature23463}
}

@article{PerreaultLevasseur2017,
  author  = {Perreault Levasseur, Laurence and Hezaveh, Yashar D. and Wechsler, Risa H.},
  title   = {Uncertainties in Parameters Estimated with Neural Networks: Application to Strong Gravitational Lensing},
  journal = {The Astrophysical Journal Letters},
  volume  = {850},
  number  = {1},
  pages   = {L7},
  year    = {2017},
  doi     = {10.3847/2041-8213/aa9704}
}

@ARTICLE{CAMELS_presentation,
author = {{Villaescusa-Navarro}, Francisco and {Angl{\'e}s-Alc{\'a}zar}, Daniel and {Genel}, Shy and {Spergel}, David N. and {Somerville}, Rachel S. and {Dave}, Romeel and {Pillepich}, Annalisa and {Hernquist}, Lars and {Nelson}, Dylan and {Torrey}, Paul and {Narayanan}, Desika and {Li}, Yin and {Philcox}, Oliver and {La Torre}, Valentina and {Maria Delgado}, Ana and {Ho}, Shirley and {Hassan}, Sultan and {Burkhart}, Blakesley and {Wadekar}, Digvijay and {Battaglia}, Nicholas and {Contardo}, Gabriella and {Bryan}, Greg L.},
title = "{The CAMELS Project: Cosmology and Astrophysics with Machine-learning Simulations}",
journal = {\apj},
year = 2021,
month = jul,
volume = {915},
number = {1},
eid = {71},
pages = {71},
doi = {10.3847/1538-4357/abf7ba},
archivePrefix = {arXiv},
eprint = {2010.00619},
primaryClass = {astro-ph.CO},
adsurl = {https://ui.adsabs.harvard.edu/abs/2021ApJ...915...71V}
}

@ARTICLE{Benyas2024ApJ,
       author = {{Benyas}, Maya and {Pfeifer}, Jordan and {Mantz}, Adam B. and {Allen}, Steven W. and {Darragh-Ford}, Elise},
        title = "{A Generative Model for Realistic Galaxy Cluster X-Ray Morphologies}",
      journal = {\apj},
         year = 2024,
        month = jul,
       volume = {969},
       number = {1},
          eid = {58},
        pages = {58},
          doi = {10.3847/1538-4357/ad5183},
archivePrefix = {arXiv},
       eprint = {2406.10456},
 primaryClass = {astro-ph.CO},
       adsurl = {https://ui.adsabs.harvard.edu/abs/2024ApJ...969...58B}
}

@ARTICLE{Wadekar2023PNAS,
       author = {{Wadekar}, Digvijay and {Thiele}, Leander and {Villaescusa-Navarro}, Francisco and {Hill}, J. Colin and {Cranmer}, Miles and {Spergel}, David N. and {Battaglia}, Nicholas and {Angl{\'e}s-Alc{\'a}zar}, Daniel and {Hernquist}, Lars and {Ho}, Shirley},
        title = "{Augmenting astrophysical scaling relations with machine learning: Application to reducing the Sunyaev-Zeldovich flux-mass scatter}",
      journal = {Proceedings of the National Academy of Science},
         year = 2023,
        month = mar,
       volume = {120},
       number = {12},
          eid = {e2202074120},
        pages = {e2202074120},
          doi = {10.1073/pnas.2202074120},
archivePrefix = {arXiv},
       eprint = {2201.01305},
 primaryClass = {astro-ph.CO},
       adsurl = {https://ui.adsabs.harvard.edu/abs/2023PNAS..12002074W}
}

@misc{Spergel2015Roman,
  author  = {Spergel, David and Gehrels, Neil and Baltay, Charles and Bennett, David and Breckinridge, James and Donahue, Megan and Dressler, Alan and others},
  title   = {Wide-Field InfrarRed Survey Telescope (WFIRST) Final Report},
  journal = {arXiv e-prints},
  year    = {2015},
  eprint  = {1503.03757},
  archivePrefix = {arXiv},
  primaryClass  = {astro-ph.IM}
}

@article{Ivezic2019LSST,
  author  = {Ivezi{\'c}, {\v{Z}}eljko and Kahn, Steven M. and Tyson, J. Anthony and Abel, Bob and Acosta, Emily and Allsman, Robyn and Alonso, David and others},
  title   = {LSST: From Science Drivers to Reference Design and Anticipated Data Products},
  journal = {Astrophysical Journal},
  volume  = {873},
  pages   = {111},
  year    = {2019},
  doi     = {10.3847/1538-4357/ab042c}
}

@article{York2000SDSS,
  author  = {York, Donald G. and Adelman, J. and Anderson, John E. and Anderson, Scott F. and Annis, James and Bahcall, Neta A. and Bakken, J. A. and Barkhouser, Robert and Bastian, Scott and Berman, Eileen and others},
  title   = {The Sloan Digital Sky Survey: Technical Summary},
  journal = {Astronomical Journal},
  volume  = {120},
  pages   = {1579--1587},
  year    = {2000},
  doi     = {10.1086/301513}
}

@misc{DESI2016,
  author  = { Levi, Michael and Bebek, Chris and Beers, Timothy and Blum, Robert and Cahn, Robert and Dawson, Kyle and Finkbeiner, Douglas and others},
  title   = {The Dark Energy Spectroscopic Instrument (DESI)},
  journal = {arXiv e-prints},
  year    = {2016},
  eprint  = {1611.00036},
  archivePrefix = {arXiv},
  primaryClass  = {astro-ph.IM}
}

@article{DESCollaboration2016,
  author  = { Abbott, T. M. C. and Abdalla, F. B. and Allam, S. and Amara, A. and Annis, J. and Asorey, J. and others},
  title   = {The Dark Energy Survey: more than dark energy -- an overview},
  journal = {Monthly Notices of the Royal Astronomical Society},
  volume  = {460},
  pages   = {1270--1299},
  year    = {2016},
  doi     = {10.1093/mnras/stw641}
}

@misc{Laureijs2011Euclid,
  author  = {Laureijs, R. and Amiaux, J. and Arduini, S. and Augu{\`e}res, J.-L. and Brinchmann, J. and Cole, R. and Cropper, M. and others},
  title   = {Euclid Definition Study Report},
  journal = {arXiv e-prints},
  year    = {2011},
  eprint  = {1110.3193},
  archivePrefix = {arXiv},
  primaryClass  = {astro-ph.CO}
}

@article{Gardner2006JWST,
  author  = {Gardner, Jonathan P. and Mather, John C. and Clampin, Mark and Doyon, Ren{\'e} and Greenhouse, Matthew A. and Hammel, Heidi B. and Hutchings, John B. and Jakobsen, Peter and Lilly, Simon J. and Long, Knox S. and others},
  title   = {The James Webb Space Telescope},
  journal = {Space Science Reviews},
  volume  = {123},
  pages   = {485--606},
  year    = {2006},
  doi     = {10.1007/s11214-006-8315-7}
}

@misc{simonyan2013,
      title={Deep Inside Convolutional Networks: Visualising Image Classification Models and Saliency Maps}, 
      author={Karen Simonyan and Andrea Vedaldi and Andrew Zisserman},
      year={2014},
      eprint={1312.6034},
      archivePrefix={arXiv},
      primaryClass={cs.CV},
      url={https://arxiv.org/abs/1312.6034}, 
}

@article{Selvaraju2017GradCAM,
  author  = {Selvaraju, Ramprasaath R. and Cogswell, Michael and Das, Abhishek and Vedantam, Ramakrishna and Parikh, Devi and Batra, Dhruv},
  title   = {Grad-CAM: Visual Explanations from Deep Networks via Gradient-Based Localization},
  journal = {International Journal of Computer Vision},
  volume  = {128},
  pages   = {336--359},
  year    = {2020},
  doi     = {10.1007/s11263-019-01228-7}
}

@misc{Mordvintsev2015,title	= {Inceptionism: Going Deeper into Neural Networks},author	= {Alexander Mordvintsev and Christopher Olah and Mike Tyka},year	= {2015},URL	= {https://research.googleblog.com/2015/06/inceptionism-going-deeper-into-neural.html}}

@ARTICLE{Rothschild2022,
       author = {{Rothschild}, Tibor and {Nagai}, Daisuke and {Aung}, Han and {Green}, Sheridan B. and {Ntampaka}, Michelle and {ZuHone}, John},
        title = "{Emulating Sunyaev-Zeldovich images of galaxy clusters using autoencoders}",
      journal = {\mnras},
         year = 2022,
        month = jun,
       volume = {513},
       number = {1},
        pages = {333-344},
          doi = {10.1093/mnras/stac438},
archivePrefix = {arXiv},
       eprint = {2110.02232},
 primaryClass = {astro-ph.CO},
       adsurl = {https://ui.adsabs.harvard.edu/abs/2022MNRAS.513..333R}
}

@ARTICLE{Caro2025,
       author = {{Caro}, Andr{\'e}s and {de Andres}, Daniel and {Cui}, Weiguang and {Yepes}, Gustavo and {De Petris}, Marco and {Ferragamo}, Antonio and {Schiltz}, F{\'e}licien and {Nef}, Am{\'e}lie},
        title = "{Deep learning generated observations of galaxy clusters from dark-matter-only simulations}",
      journal = {RAS Techniques and Instruments},
         year = 2025,
        month = jan,
       volume = {4},
          eid = {rzaf007},
        pages = {rzaf007},
          doi = {10.1093/rasti/rzaf007},
       adsurl = {https://ui.adsabs.harvard.edu/abs/2025RASTI...4....7C}
}

@ARTICLE{Srivastava2025,
       author = {{Srivastava}, Atulit and {Cui}, Weiguang and {de Andres}, Daniel and {Golden-Marx}, Jesse B. and {Rasia}, Elena and {Zu}, Ying},
        title = "{Predicting halo formation time using machine learning}",
      journal = {\aap},
         year = 2025,
        month = aug,
       volume = {700},
          eid = {A87},
        pages = {A87},
          doi = {10.1051/0004-6361/202453165},
archivePrefix = {arXiv},
       eprint = {2504.14426},
 primaryClass = {astro-ph.CO},
       adsurl = {https://ui.adsabs.harvard.edu/abs/2025A&A...700A..87S}
}

@ARTICLE{Canepa2025,
       author = {{Canepa}, Louisa and {Brough}, Sarah and {Lanusse}, Francois and {Montes}, Mireia and {Hatch}, Nina},
        title = "{Measuring the Intracluster Light Fraction with Machine Learning}",
      journal = {\apj},
         year = 2025,
        month = feb,
       volume = {980},
       number = {2},
          eid = {245},
        pages = {245},
          doi = {10.3847/1538-4357/adabc7},
archivePrefix = {arXiv},
       eprint = {2501.08378},
 primaryClass = {astro-ph.GA},
       adsurl = {https://ui.adsabs.harvard.edu/abs/2025ApJ...980..245C}
}

@ARTICLE{Marini2024,
       author = {{Marini}, I. and {Saro}, A. and {Borgani}, S. and {Boi}, M.},
        title = "{Inferring intrahalo light from stellar kinematics: A deep learning approach}",
      journal = {\aap},
         year = 2024,
        month = sep,
       volume = {689},
          eid = {A181},
        pages = {A181},
          doi = {10.1051/0004-6361/202449632},
archivePrefix = {arXiv},
       eprint = {2407.00838},
 primaryClass = {astro-ph.GA},
       adsurl = {https://ui.adsabs.harvard.edu/abs/2024A&A...689A.181M}
}

@ARTICLE{Marini2022,
       author = {{Marini}, I. and {Borgani}, S. and {Saro}, A. and {Murante}, G. and {Granato}, G.~L. and {Ragone-Figueroa}, C. and {Taffoni}, G.},
        title = "{Machine learning to identify ICL and BCG in simulated galaxy clusters}",
      journal = {\mnras},
         year = 2022,
        month = aug,
       volume = {514},
       number = {2},
        pages = {3082-3096},
          doi = {10.1093/mnras/stac1558},
archivePrefix = {arXiv},
       eprint = {2203.03360},
 primaryClass = {astro-ph.GA},
       adsurl = {https://ui.adsabs.harvard.edu/abs/2022MNRAS.514.3082M}
}

@ARTICLE{Cohn2020,
       author = {{Cohn}, J.~D. and {Battaglia}, Nicholas},
        title = "{Multiwavelength cluster mass estimates and machine learning}",
      journal = {\mnras},
         year = 2020,
        month = jan,
       volume = {491},
       number = {2},
        pages = {1575-1584},
          doi = {10.1093/mnras/stz3087},
archivePrefix = {arXiv},
       eprint = {1905.09920},
 primaryClass = {astro-ph.CO},
       adsurl = {https://ui.adsabs.harvard.edu/abs/2020MNRAS.491.1575C}
}

@ARTICLE{Iqbal2025,
       author = {{Iqbal}, Asif and {Majumdar}, Subhabrata and {Rasia}, Elena and {Pratt}, Gabriel W. and {de Andres}, Daniel and {Melin}, Jean-Baptiste and {Cui}, Weiguang},
        title = "{Deriving accurate galaxy cluster masses using X-ray thermodynamic profiles and graph neural networks}",
      journal = {\aap},
         year = 2025,
        month = dec,
       volume = {704},
          eid = {A334},
        pages = {A334},
          doi = {10.1051/0004-6361/202555691},
archivePrefix = {arXiv},
       eprint = {2509.12199},
 primaryClass = {astro-ph.CO},
       adsurl = {https://ui.adsabs.harvard.edu/abs/2025A&A...704A.334I}
}

@ARTICLE{Krippendorf2024,
       author = {{Krippendorf}, Sven and {Baron Perez}, Nicolas and {Bulbul}, Esra and {Kara}, Melih and {Seppi}, Riccardo and {Comparat}, Johan and {Artis}, Emmanuel and {Bahar}, Yunus Emre and {Garrel}, Christian and {Ghirardini}, Vittorio and {Kluge}, Matthias and {Liu}, Ang and {Ramos-Ceja}, Miriam E. and {Sanders}, Jeremy and {Zhang}, Xiaoyuan and {Br{\"u}ggen}, Marcus and {Grandis}, Sebastian and {Weller}, Jochen},
        title = "{The eROSITA Final Equatorial-Depth Survey (eFEDS): A machine learning approach to inferring galaxy cluster masses from eROSITA X-ray images}",
      journal = {\aap},
         year = 2024,
        month = feb,
       volume = {682},
          eid = {A132},
        pages = {A132},
          doi = {10.1051/0004-6361/202346826},
archivePrefix = {arXiv},
       eprint = {2305.00016},
 primaryClass = {astro-ph.CO},
       adsurl = {https://ui.adsabs.harvard.edu/abs/2024A&A...682A.132K}
}

@ARTICLE{SutherlandSDM,
       author = {{Sutherland}, Danica J. and {Xiong}, Liang and {P{\'o}czos}, Barnab{\'a}s and {Schneider}, Jeff},
        title = "{Kernels on Sample Sets via Nonparametric Divergence Estimates}",
      journal = {arXiv e-prints},
         year = 2012,
        month = feb,
          eid = {arXiv:1202.0302},
        pages = {arXiv:1202.0302},
          doi = {10.48550/arXiv.1202.0302},
archivePrefix = {arXiv},
       eprint = {1202.0302},
 primaryClass = {cs.LG},
       adsurl = {https://ui.adsabs.harvard.edu/abs/2012arXiv1202.0302S}
}

@ARTICLE{Ntampaka2016,
       author = {{Ntampaka}, M. and {Trac}, H. and {Sutherland}, D.~J. and {Fromenteau}, S. and {P{\'o}czos}, B. and {Schneider}, J.},
        title = "{Dynamical Mass Measurements of Contaminated Galaxy Clusters Using Machine Learning}",
      journal = {\apj},
         year = 2016,
        month = nov,
       volume = {831},
       number = {2},
          eid = {135},
        pages = {135},
          doi = {10.3847/0004-637X/831/2/135},
archivePrefix = {arXiv},
       eprint = {1509.05409},
 primaryClass = {astro-ph.CO},
       adsurl = {https://ui.adsabs.harvard.edu/abs/2016ApJ...831..135N}
}

@ARTICLE{Old2018,
       author = {{Old}, L. and {Wojtak}, R. and {Pearce}, F.~R. and {Gray}, M.~E. and {Mamon}, G.~A. and {Sif{\'o}n}, C. and {Tempel}, E. and {Biviano}, A. and {Yee}, H.~K.~C. and {de Carvalho}, R. and {M{\"u}ller}, V. and {Sepp}, T. and {Skibba}, R.~A. and {Croton}, D. and {Bamford}, S.~P. and {Power}, C. and {von der Linden}, A. and {Saro}, A.},
        title = "{Galaxy Cluster Mass Reconstruction Project - III. The impact of dynamical substructure on cluster mass estimates}",
      journal = {\mnras},
         year = 2018,
        month = mar,
       volume = {475},
       number = {1},
        pages = {853-866},
          doi = {10.1093/mnras/stx3241},
archivePrefix = {arXiv},
       eprint = {1709.10108},
 primaryClass = {astro-ph.CO},
       adsurl = {https://ui.adsabs.harvard.edu/abs/2018MNRAS.475..853O}
}

@ARTICLE{Ntampaka2015,
       author = {{Ntampaka}, M. and {Trac}, H. and {Sutherland}, D.~J. and {Battaglia}, N. and {P{\'o}czos}, B. and {Schneider}, J.},
        title = "{A Machine Learning Approach for Dynamical Mass Measurements of Galaxy Clusters}",
      journal = {\apj},
         year = 2015,
        month = apr,
       volume = {803},
       number = {2},
          eid = {50},
        pages = {50},
          doi = {10.1088/0004-637X/803/2/50},
archivePrefix = {arXiv},
       eprint = {1410.0686},
 primaryClass = {astro-ph.CO},
       adsurl = {https://ui.adsabs.harvard.edu/abs/2015ApJ...803...50N}
}

@ARTICLE{evrard2008virial,
       author = {{Evrard}, A.~E. and {Bialek}, J. and {Busha}, M. and {White}, M. and {Habib}, S. and {Heitmann}, K. and {Warren}, M. and {Rasia}, E. and {Tormen}, G. and {Moscardini}, L. and {Power}, C. and {Jenkins}, A.~R. and {Gao}, L. and {Frenk}, C.~S. and {Springel}, V. and {White}, S.~D.~M. and {Diemand}, J.},
        title = "{Virial Scaling of Massive Dark Matter Halos: Why Clusters Prefer a High Normalization Cosmology}",
      journal = {\apj},
         year = 2008,
        month = jan,
       volume = {672},
       number = {1},
        pages = {122-137},
          doi = {10.1086/521616},
archivePrefix = {arXiv},
       eprint = {astro-ph/0702241},
 primaryClass = {astro-ph},
       adsurl = {https://ui.adsabs.harvard.edu/abs/2008ApJ...672..122E}
}

@article{Arendt2024,
    author = {Arendt, Ashleigh R and Perrott, Yvette C and Contreras-Santos, Ana and de Andres, Daniel and Cui, Weiguang and Rennehan, Douglas},
    title = {Identifying galaxy cluster mergers with deep neural networks using idealized Compton-y and X-ray maps},
    journal = {Monthly Notices of the Royal Astronomical Society},
    volume = {530},
    number = {1},
    pages = {20-34},
    year = {2024},
    month = {05},
    issn = {0035-8711},
    doi = {10.1093/mnras/stae568},
    url = {https://doi.org/10.1093/mnras/stae568},
    eprint = {https://academic.oup.com/mnras/article-pdf/530/1/20/57170932/stae568.pdf},
}

@ARTICLE{Kaiser1986:SS,
       author = {{Kaiser}, N.},
        title = "{Evolution and clustering of rich clusters.}",
      journal = {\mnras},
         year = 1986,
        month = sep,
       volume = {222},
        pages = {323-345},
          doi = {10.1093/mnras/222.2.323},
       adsurl = {https://ui.adsabs.harvard.edu/abs/1986MNRAS.222..323K}
}

@INPROCEEDINGS{Mayet2020:Nika2,
       author = {{Mayet}, F. and {Adam}, R. and {Ade}, P. and {Andr{\'e}}, P. and {Andrianasolo}, A. and {Arnaud}, M. and {Aussel}, H. and {Bartalucci}, I. and {Beelen}, A. and {Beno{\^\i}t}, A. and {Bideaud}, A. and {Bourrion}, O. and {Calvo}, M. and {Catalano}, A. and {Comis}, B. and {De Petris}, M. and {D{\'e}sert}, F. -X. and {Doyle}, S. and {Driessen}, E.~F.~C. and {Gomez}, A. and {Goupy}, J. and {K{\'e}ruzor{\'e}}, F. and {Kramer}, C. and {Ladjelate}, B. and {Lagache}, G. and {Leclercq}, S. and {Lestrade}, J. -F. and {Mac{\'\i}as-P{\'e}rez}, J.~F. and {Mauskopf}, P. and {Monfardini}, A. and {Perotto}, L. and {Pisano}, G. and {Pointecouteau}, E. and {Ponthieu}, N. and {Pratt}, G.~W. and {Rev{\'e}ret}, V. and {Ritacco}, A. and {Romero}, C. and {Roussel}, H. and {Ruppin}, F. and {Schuster}, K. and {Shu}, S. and {Sievers}, A. and {Tucker}, C. and {Zylka}, R.},
        title = "{Cluster cosmology with the NIKA2 SZ Large Program}",
    booktitle = {mm Universe @ NIKA2 - Observing the mm Universe with the NIKA2 Camera},
         year = 2020,
       series = {European Physical Journal Web of Conferences},
       volume = {228},
        month = jun,
          eid = {00017},
        pages = {00017},
          doi = {10.1051/epjconf/202022800017},
archivePrefix = {arXiv},
       eprint = {1911.03145},
 primaryClass = {astro-ph.CO},
       adsurl = {https://ui.adsabs.harvard.edu/abs/2020EPJWC.22800017M}
}

@ARTICLE{Nclusters,
       author = {{Vikhlinin}, A. and {Kravtsov}, A.~V. and {Burenin}, R.~A. and {Ebeling}, H. and {Forman}, W.~R. and {Hornstrup}, A. and {Jones}, C. and {Murray}, S.~S. and {Nagai}, D. and {Quintana}, H. and {Voevodkin}, A.},
        title = "{Chandra Cluster Cosmology Project III: Cosmological Parameter Constraints}",
      journal = {\apj},
         year = 2009,
        month = feb,
       volume = {692},
       number = {2},
        pages = {1060-1074},
          doi = {10.1088/0004-637X/692/2/1060},
archivePrefix = {arXiv},
       eprint = {0812.2720},
 primaryClass = {astro-ph},
       adsurl = {https://ui.adsabs.harvard.edu/abs/2009ApJ...692.1060V}
}

@ARTICLE{Contreras2023ML,
       author = {{Contreras-Santos}, Ana and {Knebe}, Alexander and {Cui}, Weiguang and {Haggar}, Roan and {Pearce}, Frazer and {Gray}, Meghan and {De Petris}, Marco and {Yepes}, Gustavo},
        title = "{Galaxy pairs in THE THREE HUNDRED simulations II: studying bound ones and identifying them via machine learning}",
      journal = {\mnras},
         year = 2023,
        month = jun,
       volume = {522},
       number = {1},
        pages = {1270-1287},
          doi = {10.1093/mnras/stad1061},
archivePrefix = {arXiv},
       eprint = {2304.08898},
 primaryClass = {astro-ph.GA},
       adsurl = {https://ui.adsabs.harvard.edu/abs/2023MNRAS.522.1270C}
}

@ARTICLE{eROSITAcounts,
       author = {{Ghirardini}, V. and {Bulbul}, E. and {Artis}, E. and {Clerc}, N. and {Garrel}, C. and {Grandis}, S. and {Kluge}, M. and {Liu}, A. and {Bahar}, Y.~E. and {Balzer}, F. and {Chiu}, I. and {Comparat}, J. and {Gruen}, D. and {Kleinebreil}, F. and {Krippendorf}, S. and {Merloni}, A. and {Nandra}, K. and {Okabe}, N. and {Pacaud}, F. and {Predehl}, P. and {Ramos-Ceja}, M.~E. and {Reiprich}, T.~H. and {Sanders}, J.~S. and {Schrabback}, T. and {Seppi}, R. and {Zelmer}, S. and {Zhang}, X. and {Bornemann}, W. and {Brunner}, H. and {Burwitz}, V. and {Coutinho}, D. and {Dennerl}, K. and {Freyberg}, M. and {Friedrich}, S. and {Gaida}, R. and {Gueguen}, A. and {Haberl}, F. and {Kink}, W. and {Lamer}, G. and {Li}, X. and {Liu}, T. and {Maitra}, C. and {Meidinger}, N. and {Mueller}, S. and {Miyatake}, H. and {Miyazaki}, S. and {Robrade}, J. and {Schwope}, A. and {Stewart}, I.},
        title = "{The SRG/eROSITA all-sky survey: Cosmology constraints from cluster abundances in the western Galactic hemisphere}",
      journal = {\aap},
         year = 2024,
        month = sep,
       volume = {689},
          eid = {A298},
        pages = {A298},
          doi = {10.1051/0004-6361/202348852},
archivePrefix = {arXiv},
       eprint = {2402.08458},
 primaryClass = {astro-ph.CO},
       adsurl = {https://ui.adsabs.harvard.edu/abs/2024A&A...689A.298G}
}

@ARTICLE{Ho2022:ComaMass,
       author = {{Ho}, Matthew and {Ntampaka}, Michelle and {Rau}, Markus Michael and {Chen}, Minghan and {Lansberry}, Alexa and {Ruehle}, Faith and {Trac}, Hy},
        title = "{The dynamical mass of the Coma cluster from deep learning}",
      journal = {Nature Astronomy},
         year = 2022,
        month = jun,
       volume = {6},
        pages = {936-941},
          doi = {10.1038/s41550-022-01711-1},
archivePrefix = {arXiv},
       eprint = {2206.14834},
 primaryClass = {astro-ph.CO},
       adsurl = {https://ui.adsabs.harvard.edu/abs/2022NatAs...6..936H}
}

@ARTICLE{deAndres2024:MassMap,
       author = {{\VAN{Andres}{de Andres}{de}}, Andres, Daniel and {Cui}, Weiguang and {Yepes}, Gustavo and {De Petris}, Marco and {Ferragamo}, Antonio and {De Luca}, Federico and {Aversano}, Gianmarco and {Rennehan}, Douglas},
        title = "{The three hundred project: mapping the matter distribution in galaxy clusters via deep learning from multiview simulated observations}",
      journal = {\mnras},
         year = 2024,
        month = feb,
       volume = {528},
       number = {2},
        pages = {1517-1530},
          doi = {10.1093/mnras/stae071},
archivePrefix = {arXiv},
       eprint = {2311.02469},
 primaryClass = {astro-ph.CO},
       adsurl = {https://ui.adsabs.harvard.edu/abs/2024MNRAS.528.1517D}
}

@ARTICLE{Vogelsberger2020:reviewcosmosim,
       author = {{Vogelsberger}, Mark and {Marinacci}, Federico and {Torrey}, Paul and {Puchwein}, Ewald},
        title = "{Cosmological simulations of galaxy formation}",
      journal = {Nature Reviews Physics},
         year = 2020,
        month = jan,
       volume = {2},
       number = {1},
        pages = {42-66},
          doi = {10.1038/s42254-019-0127-2},
archivePrefix = {arXiv},
       eprint = {1909.07976},
 primaryClass = {astro-ph.GA},
       adsurl = {https://ui.adsabs.harvard.edu/abs/2020NatRP...2...42V}
}

@ARTICLE{Herbonnet2020:WL,
       author = {{Herbonnet}, Ricardo and {Sif{\'o}n}, Crist{\'o}bal and {Hoekstra}, Henk and {Bah{\'e}}, Yannick and {van der Burg}, Remco F.~J. and {Melin}, Jean-Baptiste and {von der Linden}, Anja and {Sand}, David and {Kay}, Scott and {Barnes}, David},
        title = "{CCCP and MENeaCS: (updated) weak-lensing masses for 100 galaxy clusters}",
      journal = {\mnras},
         year = 2020,
        month = oct,
       volume = {497},
       number = {4},
        pages = {4684-4703},
          doi = {10.1093/mnras/staa2303},
archivePrefix = {arXiv},
       eprint = {1912.04414},
 primaryClass = {astro-ph.CO},
       adsurl = {https://ui.adsabs.harvard.edu/abs/2020MNRAS.497.4684H}
}

@ARTICLE{Ansarifard2020,
       author = {{Ansarifard}, S. and {Rasia}, E. and {Biffi}, V. and {Borgani}, S. and {Cui}, W. and {De Petris}, M. and {Dolag}, K. and {Ettori}, S. and {Movahed}, S.~M.~S. and {Murante}, G. and {Yepes}, G.},
        title = "{The Three Hundred Project: Correcting for the hydrostatic-equilibrium mass bias in X-ray and SZ surveys}",
      journal = {\aap},
         year = 2020,
        month = feb,
       volume = {634},
          eid = {A113},
        pages = {A113},
          doi = {10.1051/0004-6361/201936742},
archivePrefix = {arXiv},
       eprint = {1911.07878},
 primaryClass = {astro-ph.CO},
       adsurl = {https://ui.adsabs.harvard.edu/abs/2020A&A...634A.113A}
}

@ARTICLE{Planck2013:SZClusterCounts,
       author = {{Planck Collaboration} and {Ade}, P.~A.~R. and {Aghanim}, N. and {Armitage-Caplan}, C. and {Arnaud}, M. and {Ashdown}, M. and {Atrio-Barandela}, F. and {Aumont}, J. and {Baccigalupi}, C. and {Banday}, A.~J. and {Barreiro}, R.~B. and {Barrena}, R. and {Bartlett}, J.~G. and {Battaner}, E. and {Battye}, R. and {Benabed}, K. and {Beno{\^\i}t}, A. and {Benoit-L{\'e}vy}, A. and {Bernard}, J. -P. and {Bersanelli}, M. and {Bielewicz}, P. and {Bikmaev}, I. and {Blanchard}, A. and {Bobin}, J. and {Bock}, J.~J. and {B{\"o}hringer}, H. and {Bonaldi}, A. and {Bond}, J.~R. and {Borrill}, J. and {Bouchet}, F.~R. and {Bourdin}, H. and {Bridges}, M. and {Brown}, M.~L. and {Bucher}, M. and {Burenin}, R. and {Burigana}, C. and {Butler}, R.~C. and {Cardoso}, J. -F. and {Carvalho}, P. and {Catalano}, A. and {Challinor}, A. and {Chamballu}, A. and {Chary}, R. -R. and {Chiang}, L. -Y. and {Chiang}, H.~C. and {Chon}, G. and {Christensen}, P.~R. and {Church}, S. and {Clements}, D.~L. and {Colombi}, S. and {Colombo}, L.~P.~L. and {Couchot}, F. and {Coulais}, A. and {Crill}, B.~P. and {Curto}, A. and {Cuttaia}, F. and {Da Silva}, A. and {Dahle}, H. and {Danese}, L. and {Davies}, R.~D. and {Davis}, R.~J. and {de Bernardis}, P. and {de Rosa}, A. and {de Zotti}, G. and {Delabrouille}, J. and {Delouis}, J. -M. and {D{\'e}mocl{\`e}s}, J. and {D{\'e}sert}, F. -X. and {Dickinson}, C. and {Diego}, J.~M. and {Dolag}, K. and {Dole}, H. and {Donzelli}, S. and {Dor{\'e}}, O. and {Douspis}, M. and {Dupac}, X. and {Efstathiou}, G. and {En{\ss}lin}, T.~A. and {Eriksen}, H.~K. and {Finelli}, F. and {Flores-Cacho}, I. and {Forni}, O. and {Frailis}, M. and {Franceschi}, E. and {Fromenteau}, S. and {Galeotta}, S. and {Ganga}, K. and {G{\'e}nova-Santos}, R.~T. and {Giard}, M. and {Giardino}, G. and {Giraud-H{\'e}raud}, Y. and {Gonz{\'a}lez-Nuevo}, J. and {G{\'o}rski}, K.~M. and {Gratton}, S. and {Gregorio}, A. and {Gruppuso}, A. and {Hansen}, F.~K. and {Hanson}, D. and {Harrison}, D. and {Henrot-Versill{\'e}}, S. and {Hern{\'a}ndez-Monteagudo}, C. and {Herranz}, D. and {Hildebrandt}, S.~R. and {Hivon}, E. and {Hobson}, M. and {Holmes}, W.~A. and {Hornstrup}, A. and {Hovest}, W. and {Huffenberger}, K.~M. and {Hurier}, G. and {Jaffe}, T.~R. and {Jaffe}, A.~H. and {Jones}, W.~C. and {Juvela}, M. and {Keih{\"a}nen}, E. and {Keskitalo}, R. and {Khamitov}, I. and {Kisner}, T.~S. and {Kneissl}, R. and {Knoche}, J. and {Knox}, L. and {Kunz}, M. and {Kurki-Suonio}, H. and {Lagache}, G. and {L{\"a}hteenm{\"a}ki}, A. and {Lamarre}, J. -M. and {Lasenby}, A. and {Laureijs}, R.~J. and {Lawrence}, C.~R. and {Leahy}, J.~P. and {Leonardi}, R. and {Le{\'o}n-Tavares}, J. and {Lesgourgues}, J. and {Liddle}, A. and {Liguori}, M. and {Lilje}, P.~B. and {Linden-V{\o}rnle}, M. and {L{\'o}pez-Caniego}, M. and {Lubin}, P.~M. and {Mac{\'\i}as-P{\'e}rez}, J.~F. and {Maffei}, B. and {Maino}, D. and {Mandolesi}, N. and {Marcos-Caballero}, A. and {Maris}, M. and {Marshall}, D.~J. and {Martin}, P.~G. and {Mart{\'\i}nez-Gonz{\'a}lez}, E. and {Masi}, S. and {Matarrese}, S. and {Matthai}, F. and {Mazzotta}, P. and {Meinhold}, P.~R. and {Melchiorri}, A. and {Melin}, J. -B. and {Mendes}, L. and {Mennella}, A. and {Migliaccio}, M. and {Mitra}, S. and {Miville-Desch{\^e}nes}, M. -A. and {Moneti}, A. and {Montier}, L. and {Morgante}, G. and {Mortlock}, D. and {Moss}, A. and {Munshi}, D. and {Naselsky}, P. and {Nati}, F. and {Natoli}, P. and {Netterfield}, C.~B. and {N{\o}rgaard-Nielsen}, H.~U. and {Noviello}, F. and {Novikov}, D. and {Novikov}, I. and {Osborne}, S. and {Oxborrow}, C.~A. and {Paci}, F. and {Pagano}, L. and {Pajot}, F. and {Paoletti}, D. and {Partridge}, B. and {Pasian}, F. and {Patanchon}, G. and {Perdereau}, O. and {Perotto}, L. and {Perrotta}, F. and {Piacentini}, F. and {Piat}, M. and {Pierpaoli}, E. and {Pietrobon}, D. and {Plaszczynski}, S. and {Pointecouteau}, E. and {Polenta}, G. and {Ponthieu}, N. and {Popa}, L. and {Poutanen}, T. and {Pratt}, G.~W. and {Pr{\'e}zeau}, G. and {Prunet}, S. and {Puget}, J. -L. and {Rachen}, J.~P. and {Rebolo}, R. and {Reinecke}, M. and {Remazeilles}, M. and {Renault}, C. and {Ricciardi}, S. and {Riller}, T. and {Ristorcelli}, I. and {Rocha}, G. and {Roman}, M. and {Rosset}, C. and {Roudier}, G. and {Rowan-Robinson}, M. and {Rubi{\~n}o-Mart{\'\i}n}, J.~A. and {Rusholme}, B. and {Sandri}, M. and {Santos}, D. and {Savini}, G. and {Scott}, D. and {Seiffert}, M.~D. and {Shellard}, E.~P.~S. and {Spencer}, L.~D. and {Starck}, J. -L. and {Stolyarov}, V. and {Stompor}, R. and {Sudiwala}, R. and {Sunyaev}, R. and {Sureau}, F. and {Sutton}, D. and {Suur-Uski}, A. -S. and {Sygnet}, J. -F. and {Tauber}, J.~A. and {Tavagnacco}, D. and {Terenzi}, L. and {Toffolatti}, L. and {Tomasi}, M. and {Tristram}, M. and {Tucci}, M. and {Tuovinen}, J. and {T{\"u}rler}, M. and {Umana}, G. and {Valenziano}, L. and {Valiviita}, J. and {Van Tent}, B. and {Vielva}, P. and {Villa}, F. and {Vittorio}, N. and {Wade}, L.~A. and {Wandelt}, B.~D. and {Weller}, J. and {White}, M. and {White}, S.~D.~M. and {Yvon}, D. and {Zacchei}, A. and {Zonca}, A.},
        title = "{Planck 2013 results. XX. Cosmology from Sunyaev-Zeldovich cluster counts}",
      journal = {\aap},
         year = 2014,
        month = nov,
       volume = {571},
          eid = {A20},
        pages = {A20},
          doi = {10.1051/0004-6361/201321521},
archivePrefix = {arXiv},
       eprint = {1303.5080},
 primaryClass = {astro-ph.CO},
       adsurl = {https://ui.adsabs.harvard.edu/abs/2014A&A...571A..20P}
}

@ARTICLE{Bohringer2012:self-similar,
       author = {{B{\"o}hringer}, H. and {Dolag}, K. and {Chon}, G.},
        title = "{Modelling self-similar appearance of galaxy clusters in X-rays}",
      journal = {\aap},
         year = 2012,
        month = mar,
       volume = {539},
          eid = {A120},
        pages = {A120},
          doi = {10.1051/0004-6361/201118000},
archivePrefix = {arXiv},
       eprint = {1112.5035},
 primaryClass = {astro-ph.CO},
       adsurl = {https://ui.adsabs.harvard.edu/abs/2012A&A...539A.120B}
}

@incollection{Lovisari2022:scalinglaws,
  author    = {Lovisari, Lorenzo and Maughan, Ben J.},
  title     = {Scaling Relations of Clusters and Groups and Their Evolution},
  booktitle = {Handbook of X-ray and Gamma-ray Astrophysics},
  editor    = {Bambi, Cosimo and Santangelo, Andrea},
  publisher = {Springer},
  address   = {Singapore},
  year      = {2022},
  doi       = {10.1007/978-981-16-4544-0_118-1}
}

@ARTICLE{NFW,
       author = {{Navarro}, Julio F. and {Frenk}, Carlos S. and {White}, Simon D.~M.},
        title = "{Simulations of X-ray clusters}",
      journal = {\mnras},
         year = 1995,
        month = aug,
       volume = {275},
       number = {3},
        pages = {720-740},
          doi = {10.1093/mnras/275.3.720},
archivePrefix = {arXiv},
       eprint = {astro-ph/9408069},
 primaryClass = {astro-ph},
       adsurl = {https://ui.adsabs.harvard.edu/abs/1995MNRAS.275..720N}
}

@ARTICLE{Zwicky1933,
       author = {{Zwicky}, F.},
        title = "{Die Rotverschiebung von extragalaktischen Nebeln}",
      journal = {Helvetica Physica Acta},
         year = 1933,
        month = jan,
       volume = {6},
        pages = {110-127},
       adsurl = {https://ui.adsabs.harvard.edu/abs/1933AcHPh...6..110Z}
}

@ARTICLE{Cui2022,
       author = {{Cui}, Weiguang and {Dave}, Romeel and {Knebe}, Alexander and {Rasia}, Elena and {Gray}, Meghan and {Pearce}, Frazer and {Power}, Chris and {Yepes}, Gustavo and {Anbajagane}, Dhayaa and {Ceverino}, Daniel and {Contreras-Santos}, Ana and {de Andres}, Daniel and {De Petris}, Marco and {Ettori}, Stefano and {Haggar}, Roan and {Li}, Qingyang and {Wang}, Yang and {Yang}, Xiaohu and {Borgani}, Stefano and {Dolag}, Klaus and {Zu}, Ying and {Kuchner}, Ulrike and {Ca{\~n}as}, Rodrigo and {Ferragamo}, Antonio and {Gianfagna}, Giulia},
        title = "{THE THREE HUNDRED project: The GIZMO-SIMBA run}",
      journal = {\mnras},
         year = 2022,
        month = jul,
       volume = {514},
       number = {1},
        pages = {977-996},
          doi = {10.1093/mnras/stac1402},
archivePrefix = {arXiv},
       eprint = {2202.14038},
 primaryClass = {astro-ph.GA},
       adsurl = {https://ui.adsabs.harvard.edu/abs/2022MNRAS.514..977C}
}

@ARTICLE{Ho2023,
       author = {{Ho}, Matthew and {Soltis}, John and {Farahi}, Arya and {Nagai}, Daisuke and {Evrard}, August and {Ntampaka}, Michelle},
        title = "{Benchmarks and explanations for deep learning estimates of X-ray galaxy cluster masses}",
      journal = {\mnras},
         year = 2023,
        month = sep,
       volume = {524},
       number = {3},
        pages = {3289-3302},
          doi = {10.1093/mnras/stad2005},
archivePrefix = {arXiv},
       eprint = {2303.00005},
 primaryClass = {astro-ph.CO},
       adsurl = {https://ui.adsabs.harvard.edu/abs/2023MNRAS.524.3289H}
}

@ARTICLE{Kodi2020,
       author = {{Kodi Ramanah}, Doogesh and {Wojtak}, Rados{\l}aw and {Ansari}, Zoe and {Gall}, Christa and {Hjorth}, Jens},
        title = "{Dynamical mass inference of galaxy clusters with neural flows}",
      journal = {\mnras},
         year = 2020,
        month = dec,
       volume = {499},
       number = {2},
        pages = {1985-1997},
          doi = {10.1093/mnras/staa2886},
archivePrefix = {arXiv},
       eprint = {2003.05951},
 primaryClass = {astro-ph.CO},
       adsurl = {https://ui.adsabs.harvard.edu/abs/2020MNRAS.499.1985K}
}

@ARTICLE{Ferragamo2023,
       author = {{Ferragamo}, A. and {de Andres}, D. and {Sbriglio}, A. and {Cui}, W. and {De Petris}, M. and {Yepes}, G. and {Dupuis}, R. and {Jarraya}, M. and {Lahouli}, I. and {De Luca}, F. and {Gianfagna}, G. and {Rasia}, E.},
        title = "{THE THREE HUNDRED project: a machine learning method to infer clusters of galaxy mass radial profiles from mock Sunyaev-Zel'dovich maps}",
      journal = {\mnras},
         year = 2023,
        month = apr,
       volume = {520},
       number = {3},
        pages = {4000-4008},
          doi = {10.1093/mnras/stad377},
archivePrefix = {arXiv},
       eprint = {2207.12337},
 primaryClass = {astro-ph.CO},
       adsurl = {https://ui.adsabs.harvard.edu/abs/2023MNRAS.520.4000F}
}

@ARTICLE{exmachina2023,
       author = {{Smith}, Michael J. and {Geach}, James E.},
        title = "{Astronomia ex machina: a history, primer and outlook on neural networks in astronomy}",
      journal = {Royal Society Open Science},
         year = 2023,
        month = may,
       volume = {10},
       number = {5},
          eid = {221454},
        pages = {221454},
          doi = {10.1098/rsos.221454},
archivePrefix = {arXiv},
       eprint = {2211.03796},
 primaryClass = {astro-ph.IM},
       adsurl = {https://ui.adsabs.harvard.edu/abs/2023RSOS...1021454S}
}

@ARTICLE{MDPL2,
       author = {{Klypin}, Anatoly and {Yepes}, Gustavo and {Gottl{\"o}ber}, Stefan and {Prada}, Francisco and {He{\ss}}, Steffen},
        title = "{MultiDark simulations: the story of dark matter halo concentrations and density profiles}",
      journal = {\mnras},
         year = 2016,
        month = apr,
       volume = {457},
       number = {4},
        pages = {4340-4359},
          doi = {10.1093/mnras/stw248},
archivePrefix = {arXiv},
       eprint = {1411.4001},
 primaryClass = {astro-ph.CO},
       adsurl = {https://ui.adsabs.harvard.edu/abs/2016MNRAS.457.4340K}
}

@ARTICLE{deAndres2022Planck,
       author = {{\VAN{Andres}{de Andres}{de}}, Andres, Daniel and {Cui}, Weiguang and {Ruppin}, Florian and {De Petris}, Marco and {Yepes}, Gustavo and {Gianfagna}, Giulia and {Lahouli}, Ichraf and {Aversano}, Gianmarco and {Dupuis}, Romain and {Jarraya}, Mahmoud and {Vega-Ferrero}, Jes{\'u}s},
        title = "{A deep learning approach to infer galaxy cluster masses from Planck Compton-y parameter maps}",
      journal = {Nature Astronomy},
         year = 2022,
        month = nov,
       volume = {6},
        pages = {1325-1331},
          doi = {10.1038/s41550-022-01784-y},
archivePrefix = {arXiv},
       eprint = {2209.10333},
 primaryClass = {astro-ph.CO},
       adsurl = {https://ui.adsabs.harvard.edu/abs/2022NatAs...6.1325D}
}

@ARTICLE{deandres2022baryon,
       author = {{\VAN{Andres}{de Andres}{de}}, Andres, Daniel and {Yepes}, Gustavo and {Sembolini}, Federico and {Mart{\'\i}nez-Mu{\~n}oz}, Gonzalo and {Cui}, Weiguang and {Robledo}, Francisco and {Chuang}, Chia-Hsun and {Rasia}, Elena},
        title = "{Machine learning methods to estimate observational properties of galaxy clusters in large volume cosmological N-body simulations}",
      journal = {\mnras},
         year = 2023,
        month = jan,
       volume = {518},
       number = {1},
        pages = {111-129},
          doi = {10.1093/mnras/stac3009},
archivePrefix = {arXiv},
       eprint = {2204.10751},
 primaryClass = {astro-ph.CO},
       adsurl = {https://ui.adsabs.harvard.edu/abs/2023MNRAS.518..111D}
}

@ARTICLE{Gianfagna2023,
       author = {{Gianfagna}, Giulia and {Rasia}, Elena and {Cui}, Weiguang and {De Petris}, Marco and {Yepes}, Gustavo and {Contreras-Santos}, Ana and {Knebe}, Alexander},
        title = "{A study of the hydrostatic mass bias dependence and evolution within The Three Hundred clusters}",
      journal = {\mnras},
         year = 2023,
        month = jan,
       volume = {518},
       number = {3},
        pages = {4238-4248},
          doi = {10.1093/mnras/stac3364},
archivePrefix = {arXiv},
       eprint = {2211.08372},
 primaryClass = {astro-ph.CO},
       adsurl = {https://ui.adsabs.harvard.edu/abs/2023MNRAS.518.4238G}
}

@ARTICLE{Cui2018,
       author = {{Cui}, Weiguang and {Knebe}, Alexander and {Yepes}, Gustavo and {Pearce}, Frazer and {Power}, Chris and {Dave}, Romeel and {Arth}, Alexander and {Borgani}, Stefano and {Dolag}, Klaus and {Elahi}, Pascal and {Mostoghiu}, Robert and {Murante}, Giuseppe and {Rasia}, Elena and {Stoppacher}, Doris and {Vega-Ferrero}, Jesus and {Wang}, Yang and {Yang}, Xiaohu and {Benson}, Andrew and {Cora}, Sof{\'\i}a A. and {Croton}, Darren J. and {Sinha}, Manodeep and {Stevens}, Adam R.~H. and {Vega-Mart{\'\i}nez}, Cristian A. and {Arthur}, Jake and {Baldi}, Anna S. and {Ca{\~n}as}, Rodrigo and {Cialone}, Giammarco and {Cunnama}, Daniel and {De Petris}, Marco and {Durando}, Giacomo and {Ettori}, Stefano and {Gottl{\"o}ber}, Stefan and {Nuza}, Sebasti{\'a}n E. and {Old}, Lyndsay J. and {Pilipenko}, Sergey and {Sorce}, Jenny G. and {Welker}, Charlotte},
        title = "{The Three Hundred project: a large catalogue of theoretically modelled galaxy clusters for cosmological and astrophysical applications}",
      journal = {\mnras},
         year = 2018,
        month = nov,
       volume = {480},
       number = {3},
        pages = {2898-2915},
          doi = {10.1093/mnras/sty2111},
archivePrefix = {arXiv},
       eprint = {1809.04622},
 primaryClass = {astro-ph.GA},
       adsurl = {https://ui.adsabs.harvard.edu/abs/2018MNRAS.480.2898C}
}

@ARTICLE{Ho2021,
       author = {{Ho}, Matthew and {Farahi}, Arya and {Rau}, Markus Michael and {Trac}, Hy},
        title = "{Approximate Bayesian Uncertainties on Deep Learning Dynamical Mass Estimates of Galaxy Clusters}",
      journal = {\apj},
         year = 2021,
        month = feb,
       volume = {908},
       number = {2},
          eid = {204},
        pages = {204},
          doi = {10.3847/1538-4357/abd101},
archivePrefix = {arXiv},
       eprint = {2006.13231},
 primaryClass = {astro-ph.CO},
       adsurl = {https://ui.adsabs.harvard.edu/abs/2021ApJ...908..204H}
}

@ARTICLE{Kodi2021,
       author = {{Kodi Ramanah}, Doogesh and {Wojtak}, Rados{\l}aw and {Arendse}, Nikki},
        title = "{Simulation-based inference of dynamical galaxy cluster masses with 3D convolutional neural networks}",
      journal = {\mnras},
         year = 2021,
        month = mar,
       volume = {501},
       number = {3},
        pages = {4080-4091},
          doi = {10.1093/mnras/staa3922},
archivePrefix = {arXiv},
       eprint = {2009.03340},
 primaryClass = {astro-ph.CO},
       adsurl = {https://ui.adsabs.harvard.edu/abs/2021MNRAS.501.4080K}
}

@ARTICLE{Ho2019,
       author = {{Ho}, Matthew and {Rau}, Markus Michael and {Ntampaka}, Michelle and {Farahi}, Arya and {Trac}, Hy and {P{\'o}czos}, Barnab{\'a}s},
        title = "{A Robust and Efficient Deep Learning Method for Dynamical Mass Measurements of Galaxy Clusters}",
      journal = {\apj},
         year = 2019,
        month = dec,
       volume = {887},
       number = {1},
          eid = {25},
        pages = {25},
          doi = {10.3847/1538-4357/ab4f82},
archivePrefix = {arXiv},
       eprint = {1902.05950},
 primaryClass = {astro-ph.CO},
       adsurl = {https://ui.adsabs.harvard.edu/abs/2019ApJ...887...25H}
}

@ARTICLE{Yan2020,
       author = {{Yan}, Z. and {Mead}, A.~J. and {Van Waerbeke}, L. and {Hinshaw}, G. and {McCarthy}, I.~G.},
        title = "{Galaxy cluster mass estimation with deep learning and hydrodynamical simulations}",
      journal = {\mnras},
         year = 2020,
        month = dec,
       volume = {499},
       number = {3},
        pages = {3445-3458},
          doi = {10.1093/mnras/staa3030},
archivePrefix = {arXiv},
       eprint = {2005.11819},
 primaryClass = {astro-ph.CO},
       adsurl = {https://ui.adsabs.harvard.edu/abs/2020MNRAS.499.3445Y}
}

@ARTICLE{Gupta2020,
       author = {{Gupta}, N. and {Reichardt}, C.~L.},
        title = "{Mass Estimation of Galaxy Clusters with Deep Learning. I. Sunyaev-Zel'dovich Effect}",
      journal = {\apj},
         year = 2020,
        month = sep,
       volume = {900},
       number = {2},
          eid = {110},
        pages = {110},
          doi = {10.3847/1538-4357/aba694},
archivePrefix = {arXiv},
       eprint = {2003.06135},
 primaryClass = {astro-ph.CO},
       adsurl = {https://ui.adsabs.harvard.edu/abs/2020ApJ...900..110G}
}

@ARTICLE{Ntampaka2019,
       author = {{Ntampaka}, M. and {ZuHone}, J. and {Eisenstein}, D. and {Nagai}, D. and {Vikhlinin}, A. and {Hernquist}, L. and {Marinacci}, F. and {Nelson}, D. and {Pakmor}, R. and {Pillepich}, A. and {Torrey}, P. and {Vogelsberger}, M.},
        title = "{A Deep Learning Approach to Galaxy Cluster X-Ray Masses}",
      journal = {\apj},
         year = 2019,
        month = may,
       volume = {876},
       number = {1},
          eid = {82},
        pages = {82},
          doi = {10.3847/1538-4357/ab14eb},
archivePrefix = {arXiv},
       eprint = {1810.07703},
 primaryClass = {astro-ph.CO},
       adsurl = {https://ui.adsabs.harvard.edu/abs/2019ApJ...876...82N}
}

@ARTICLE{Huertas2022,
       author = {{Huertas-Company}, M. and {Lanusse}, F.},
        title = "{The Dawes Review 10: The impact of deep learning for the analysis of galaxy surveys}",
      journal = {\pasa},
         year = 2023,
        month = jan,
       volume = {40},
          eid = {e001},
        pages = {e001},
          doi = {10.1017/pasa.2022.55},
archivePrefix = {arXiv},
       eprint = {2210.01813},
 primaryClass = {astro-ph.IM},
       adsurl = {https://ui.adsabs.harvard.edu/abs/2023PASA...40....1H}
}


\end{document}